\documentclass[journal,draftcls,onecolumn,12pt,twoside]{IEEEtran}

\usepackage{enumitem}
\usepackage{amstext}
\usepackage{graphicx}
\usepackage{epstopdf}
\usepackage{amsmath, amsthm, amssymb, amsfonts,bm,breqn}
\usepackage{hyperref}
\usepackage{stfloats}
\usepackage[caption=false,font=footnotesize]{subfig}
\usepackage{csquotes}
\usepackage{flushend}

\usepackage{lipsum}
\makeatletter
\newtheoremstyle{mytheorem}
  {3pt}
  {3pt}
  {\itshape}
  {}
  {\itshape \textsf}
  {.}
  {.5em}
  {\thmname{#1}\thmnumber{\@ifnotempty{#1}{ }#2}%
   \thmnote{ {\the\thm@notefont(#3)}}}
\makeatother
\theoremstyle{mytheorem}
\newtheorem{thm}{Theorem}
\newtheorem{prop}[thm]{Proposition}

\newtheorem{cor}[thm]{Corollary}
\newtheorem{lem}[thm]{Lemma}

\newtheorem{remark}{Remark}

\begin{document}
\title{The Impact of Physical Channel on Performance of Subspace-Based Channel Estimation in Massive MIMO Systems}
\author{Mohammed~Teeti,~\IEEEmembership{Member,~IEEE,}
Jun~Sun, David~Gesbert,~\IEEEmembership{Fellow,~IEEE,} and~
YingZhuang Liu%
\thanks{M. Teeti, J. Sun, and Y. Liu are with the Department of Electronics and Information Engineering, Huazhong University of Science \& Technology, Wuhan, 430074, China (e-mail: teeti.moh@gmail.com, francissunj@gmail.com and liuyz@mail.hust.edu.cn).}%
\thanks{D. Gesbert is with the Mobile Communications Department of EURECOM, EURECOM, 06410 Biot, France (e-mail:  david.gesbert@eurecom.fr).}}%

\maketitle
\begin{abstract}
A subspace method for channel estimation has been recently proposed~\cite{analysis_blind_pilot} for tackling the pilot contamination effect, which is regarded  by some researchers  as a bottleneck in massive MIMO systems. It was shown in~\cite{analysis_blind_pilot} that if the power ratio between the desired signal and interference is kept above a certain value, the received signal spectrum splits into signal and interference eigenvalues, namely, the \enquote{pilot contamination} effect can be completely eliminated. However,~\cite{analysis_blind_pilot} assumes an independently distributed (i.d.) channel, which is actually not much the case in practice. Considering this, a more sensible finite-dimensional physical channel model (i.e., a finite scattering environment, where signals impinge on the base station (BS) from a finite number of angles of arrival (AoA)) is employed in this paper. Via asymptotic spectral analysis, it is demonstrated that, compared with the i.d. channel, the physical channel imposes a penalty in the form of an increased power ratio between the useful signal and the interference. Furthermore, we demonstrate an interesting \enquote{antenna saturation} effect, i.e., when the number of the BS antennas approaches infinity, the performance under the physical channel with $P$ AoAs is limited by and nearly the same as the performance under the i.d. channel with $P$ receive antennas.
\end{abstract}

\begin{IEEEkeywords}
massive MIMO, physical channel, subspace method, random matrix theory, asymptotic eigenvalue distribution
\end{IEEEkeywords}

\IEEEpeerreviewmaketitle

\let\incfile\include
\renewcommand{\include}[1]{\directlua{tex.ConvertToSpace("#1")}}

\section{Introduction}\label{sec1}
\IEEEPARstart{M}{assive} MIMO systems that employ a large number of antennas at the base station (BS) have attracted significant interest recently~\cite{Marzetta2010},~\cite{6241389},~\cite{6375940},~\cite{spectral_energy_eff}. The main advantage of using massive MIMO lies in the significant improvement of spectral and energy efficiency. However, when we shift our attention to the multicell scenario, the massive MIMO system would, unfortunately, be plagued by the so-called \enquote{pilot contamination} effect, which is due to the use of non-fully orthogonal pilot sequences across all the cells. However, it is often hard to achieve full orthogonality of the pilot sequences among the terminals across the cells, due to the limited coherence time of the mobile communication channels. The performance of massive MIMO systems under the full reuse of pilot sequences was studied intensively by Marzetta~\cite{Marzetta2010}, where it is shown that, by linear processing of the received signal at the BS, the performance is only interference-limited and does not depend on the transmitted power of users, i.e., signal-to-interference ratio (SIR) cannot grow unboundedly.

Several motivating works have been conducted aiming to address the above pilot contamination problem. In~\cite{6415397}, authors use a coordinated channel estimation scheme which exploits the information embedded in the second-order statistics (e.g., covariance matrix) of uplink channels. The main idea of the scheme in~\cite{6415397} is to assign pilots to users associated with covariance matrices exhibiting maximum  orthogonality of signal subspaces. Although this scheme greatly alleviates the pilot contamination, but it incurs too much coordination since all covariance matrices of all uplink channels must be learned beforehand.

In~\cite{EVD}, by leveraging a key observation that the quasi orthogonality among the channel vectors implies that the channel vectors of the desired users are eigenvectors of the covariance matrix of the received signal in the asymptotic limit, Ngo and Larsson~\cite{EVD} proposed a blind channel estimation method which mitigates the need of pilots. However, it is worth noting this scheme heavily hinges on a large number of BS antennas as well as large block length (i.e., sample size), therefore its performance might be unsatisfactory in the \enquote{not very large} regime.

M$\ddot{\mbox{u}}$ller et. al.~\cite{blind_pilot_cont} proposed another blind pilot decontamination method which, in essence, aims to distinguish users in the amplitude domain by exploiting the difference between the channel gains of the intended users and the interfering users. Therefore, this work can be regarded as a parallel to~\cite{6415397}, which essentially attempts to distinguish users with the same pilot sequence in the angular domain by exploiting the non-isotropy of angles of arrival (AoA) multipaths of the channel. By means of random matrix theory (RMT) and free probability theory (FPT),~\cite{blind_pilot_cont} demonstrates that under a mild power ratio the subspace-based estimation scheme is capable of totally removing the interference and hence completely tackling the pilot contamination problem.

We note that the channel considered in~\cite{blind_pilot_cont} consists of independently distributed (i.d.) Rayleigh-faded entries, implicitly implying a rich scattering environment. However, in the real world, it is quite common that the number of scatterers is limited and correspondingly the number of AoAs is finite~\cite{Muller:2006:RMM:2263243.2267030},~\cite{6415397},~\cite{6493983}. Therefore, considering a more sensible physical channel model (or alternatively, called finite-dimensional channel, due to the finiteness of its degrees of freedom) is of significant importance since the finiteness of degrees of freedom of the channel is relevant to the ability of subspace method to identify the eigenvalues of desired and interfering users. In the absence of a priori knowledge of the channel at the BS, it is natural to assume the AoAs in the channel model to be uniformly distributed in the interval $[0,\pi]$. Therefore, the total number of AoAs, which roughly corresponds to the total number of scatterers around the BS, will be the only key parameter (besides the ratio of the channel gains that correspond to the desired users and the interfering users, respectively) that might have crucial impact on the performance of the blind subspace method. Hence a natural question one may ask: What is the impact of finite-dimensional channel on the performance of the subspace-based channel estimation scheme? What is the price paid for this finite dimensionality of the channel? Moreover, does increasing the number of antennas at the BS help to alleviate the performance degradation?

\begin{figure}[!ht]
\centering
\includegraphics[width=2.7in]{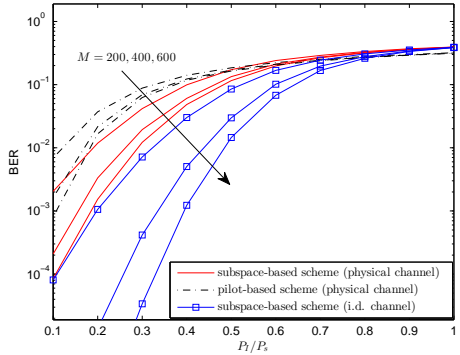}
\caption{ Antenna saturation effect under physical channel. $L=4$ cells, $K=5$ users per cell, $P=200$ AoAs, coherence time $N=400$ symbol periods and per-user $\mathrm{SNR} = -5\mathrm{dB}$. ZF is used for channel estimation and MF for data detection. Array elements are critically-spaced.}
\label{fig7}
\end{figure}

To answer the above questions, the core task is to characterize the spectrum of the observed signal under the physical channel model. Despite its difficulty, we manage that by leveraging tools in RMT and FPT. Intuitively, the case of the i.d. channel can be regarded as a special case of our result, i.e., when the number of AoAs approaches infinity. The contributions of this paper are summarized as follows:
\begin{enumerate}[leftmargin=*]
\item It is shown that to guarantee the performance of the subspace-based channel estimation scheme under the physical channel, we should pay a cost of an extra power margin between the intended users and the interfering users with respect to (w.r.t.) a given BS.
    \item For multicell multiuser MIMO system where the users of each cell are seen from distinct AoAs w.r.t. a given BS, it is shown that the performance is mainly determined by the cell associated with smaller number of AoAs.

\item  It is demonstrated that there exists an \emph{antenna saturation} effect at the BS, i.e. adding antennas beyond a threshold at the BS is of limited help to enhance the performance under the physical channel model. To be concrete, the performance under a physical channel with $P$ AoAs is nearly the same as the performance under the i.d channel model with $P$ antennas at the BS (see Fig.\ref{fig7}, and Sec.\ref{sec5} for a detailed description of the simulation setup and discussion).
\end{enumerate}
 \begin{figure}[!ht]
\centering
\includegraphics[width=2.7in]{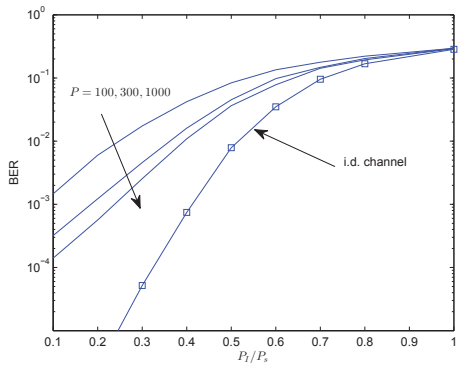}
\caption{The effect of the number of AoAs on the required power ratio $P_I/P_s$. $M=200$ receive antennas, $L=2$ cells, $K=5$ users per cell, coherence time $N=400$ symbol periods, per-user $\mathrm{SNR}=0 \mathrm{dB}$. ZF is used for channel estimation and MF for data detection. Array elements are critically-spaced.}
\label{IllustrativeFigure}
\end{figure}

In the following we provide some intuitive remarks of the above results.
When the number of AoAs is decreased, the correlation (or more exactly, the coherence) among the channel vectors will increase correspondingly. As a result, the condition number (or eigenvalue spread) of the channel matrix $\bm{H}_s$ (corresponding to inside-cell users) as well as $\bm{H}_I$ (corresponding to outside-cell users) will both increase. Therefore, the left and right endpoints of the eigenvalue clusters corresponding to $\bm{H}_s$ and $\bm{H}_I$, respectively, tend to be closer, even overlapping (e.g., see Fig.\ref{fig3}). In order to keep the above two eigenvalue clusters apart, we have to pay a cost of boosting the channel gain ratio between the desired users and the interfering users. Fig.\ref{IllustrativeFigure} illustrates the BER performance of the subspace method proposed in~\cite{blind_pilot_cont} when we vary the number of AoAs, denoted $P$. It can be seen from Fig.\ref{IllustrativeFigure} that the smaller $P$ is, the higher channel gain ratio ${P_s}/{P_I}$ is required.

As a final remark, it is worthwhile of pointing out the different role that channel correlation plays in the two pilot decontamination methods, namely, the Bayesian channel estimation in~\cite{6415397} and the subspace method in~\cite{blind_pilot_cont} as well as this paper, that is, while the correlation is beneficial in the former method, it is however, disadvantageous in the latter. The key reason for this difference lies in that, in a nutshell, the former is a \emph{Bayesian} method while the latter is a \emph{non-Bayesian} one (i.e., no a priori assumed). Specifically, the correlation is advantageous for the \emph{linear} (MMSE) channel estimation (since it helps to distinguish users in the angular domain) at the cost of acquiring the a priori (namely, the covariance matrices) and coordinating the pilots. On the other hand, as a non-Bayesian method in essence, the subspace method performs the channel estimation in a \emph{nonlinear} way, which highly relies on the \emph{instantaneous} property (such as the eigenvalue spread of the channel matrix), rather than the \emph{statistical} property in the Bayesian method~\cite{6415397}. Since correlation tends to increase the eigenvalue spread, it will impact the ability of subspace method to distinguish the desired users and interfering users via eigenvalues clustering.

The rest of the paper is organized as follows:
In Sec.\ref{sec2} we introduce the physical channel model and review the subspace-based channel estimation. In Sec.\ref{sec4}, the asymptotic eigenvalue distribution (AED) of the channel is derived in the Stieltjes domain. This will enable us to derive analytical expressions, from which the support of the distribution can be identified, hence yielding the main results of this paper.  Further, Sec.\ref{sec5} leverages the obtained formulae to demonstrate the impact of the physical channel on the spectral spread of signal and interference subspaces, and BER simulation results are presented for performance comparison. Sec.\ref{sec6} summarizes the main results and concludes this paper.

\section{problem formulation} \label{sec2}

\subsection{Channel model}
We will consider the uplink in multicell multiuser MIMO communication system with $L$ cells. Each BS is equipped with a uniform linear array with a large number of antennas $M$, serving $K$ single-antenna users. We also assume time-division duplex architecture. The channel is assumed narrowband flat-fading, which remains constant over a coherence time of $N$ symbol periods, and changes independently from one coherence time to another (i.e., block fading channel~\cite{Marzetta99}). Furthermore, we assume time-synchronous users, and full reuse of pilot sequences among all cells to facilitate channel estimation at the BS.

Since the channel is estimated blindly, we assume no a priori information about the channel is available at BSs, including the angular spread, which is in contrast to the Bayesian channel estimation method~\cite{6415397}. Thus, we assume all AoAs are uniformly distributed in $ [0, \pi]$, which is a reasonable, yet mathematically tractable assumption. However, the theoretical analysis in our paper gives insights into the performance of the subspace method under a physical channel with small angular spreads as well.

The array response in a given direction is quantified by the so-called \emph{steering vector}. Hence, the channel from the $\emph{k}$th user in the $\emph{i}$th  cell to the intended BS, denoted ${{\bm{h}}_{ki}}$, can be modeled as a linear combination of all steering vectors (see e.g.,~\cite{6493983},~\cite{6415397})
\begin{eqnarray}\label{eq:steering}
{{\bm{h}}_{ki}} = \frac{1}{\sqrt{P_{ki}}}\sum\limits_{j = 1}^{P_{ki}} {{\alpha _{kij}}{\bm{s}}({\varphi _{kij}})}
\end{eqnarray}
where $P_{ki}$ is the number of i.i.d. AoA multipaths, ${\bm{s}}({\varphi _{kij}})\in \mathcal{C}^{M}$ is the steering vector corresponding to the angle of arrival $\varphi _{kij}$ associated with the $\emph{j}$th path, $1 / \sqrt{P_{ki}}$ serves as a normalization factor and $\alpha _{kij}\sim\mathcal{CN}(0,\beta_{ki})$ denotes the channel gain associated with the \emph{j}th direction, where $\sqrt{\beta_{ki}}$ is the average channel attenuation, due to path loss and shadowing effect. Moreover, it is assumed that all $\varphi_{kij}$ and $\alpha_{kij}$ are independent over user index \emph{k}, cell index \emph{i} and direction index \emph{j}. The length-$M$ steering vector associated with the angle of arrival $\varphi _{kij}$ is given by ${\bm{s}}({\varphi _{kij}})= \left ( e^{-\bm{j} f_1(\varphi _{kij})}, e^{-\bm{j} f_2(\varphi _{kij})}, \cdots, e^{-\bm{j}f_M(\varphi _{kij})}\right )^T$,
\begin{figure}
\centering
\includegraphics[width=2.8in]{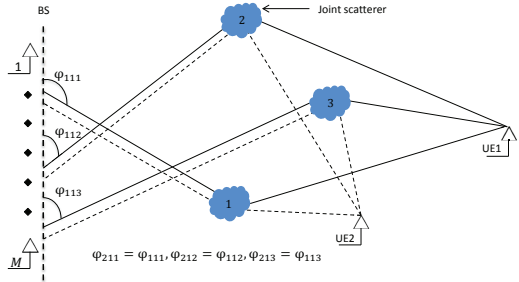}
\caption{A simple illustration of multipath environment where the signals of users 1 and 2 arrive at the BS from identical AoAs due to joint scatterers.}\label{ccccccccc}
\end{figure}
where we use \enquote{$\bm{j}$} to denote the imaginary unit and $f_i(\varphi _{kij})$ is a function of $\varphi _{kij}$.

Let
\begin{equation}\label{eq:physical channel model}
{\bm{H}}_{i} \stackrel{\Delta}{=}  ( \bm {S}_{1i} \tilde{\bm{h}}_{1i}, \cdots,  \bm {S}_{Ki} \tilde{\bm{h}}_{Ki} )
\end{equation}
be the $M\times K$ effective fast-fading channel from $K$ users in the \emph{i}th cell to the intended BS, where $\bm{S}_{ki} =( {\bm{s}}({\varphi _{ki1}}), \cdots,{\bm{s}}({\varphi _{kiP_{ki}}}) )/\sqrt{P_{ki}} \in \mathcal{C}^{M \times P_{ki}}$ comprises all steering vectors of the $\emph{k}$th user in the $i$th cell w.r.t. the intended BS, and $\tilde{\bm{h}}_{ki} \in \mathcal{C}^{P_{ki}\times 1}$ consists of i.i.d. $\mathcal{CN}(0,1)$ entries. The $M\times N$ signal received by the $\emph{l}$th BS during $N$ consecutive symbol intervals may be written as
\begin{eqnarray}\label{eq: system model 12}
\bm{Y}_l = \sum\limits_{i = 1}^L {\bm{H}}_{i}\bm{D}_{i}^{1/2}{\bm{T}_i}^{1/2} {\bm{X}_i} + \bm{W}_l
\end{eqnarray}
where $\bm{D}_{i}=\operatorname{diag}({\beta_{1i}},{\beta_{2i}},\cdots,{\beta_{Ki}})\in \mathcal{R}^{K\times K}$ comprises large-scale fading coefficients, ${\bm{T}_i}=\operatorname{diag}(p_{1i},\cdots,p_{Ki})\in \mathcal{R}^{K\times K}$ consists of the average transmission powers, and $\bm{X}_i \in \mathcal{C}^{K \times N}$ is the normalized input symbols (preceded by short pilot signals), all for $K$ users in the $\emph{i}$th cell. The entries of the noise matrix $\bm{W}_l$ are assumed i.i.d. $\mathcal{CN}(0,1)$.

For the sake of analytical simplicity, we shall assume that all users belonging to the same (say, \emph{i}th) cell are seen from the same set of directions (whose cardinality is $P_i$, and a corresponding steering matrix $\bm{S}_i$) w.r.t. the intended BS. Therefore, in this \emph{distinct} AoAs scenario,~\eqref{eq:physical channel model} takes the form of ${\bm{H}}_{i}=\bm{S}_i\tilde{\bm{H}_i}$, where $\tilde{\bm{H}_i} \in \mathcal{C}^{P_i \times K}$ consists of i.i.d. $\mathcal{CN}(0,1)$ entries, correspondingly,~\eqref{eq: system model 12} can now be rewritten as:
\begin{equation}\label{eq: system model 13}
\bm{Y}_l = \sum\limits_{i = 1}^L {\bm{S}}_{i} \tilde{\bm{H}}_{i} \bm{D}_{i}^{1/2}{\bm{T}_i}^{1/2} {\bm{X}_i} + \bm{W}_l.
\end{equation}

As a first step in addressing the channel model~\eqref{eq: system model 13}, we'll make a further assumption that all $\bm{S}_i$ are identical w.r.t. to the intended BS, which means all users in the network are seen from the same set of $P$ AoAs by the intended BS (for the sake of illustration, see Fig.\ref{ccccccccc}). Hence, in this \emph{identical} AoAs scenario,~\eqref{eq: system model 13} can take a simpler form as:
\begin{equation}\label{eq: system model 1}
\bm{Y}_l= \bm {S} \sum\limits_{i = 1}^L  \tilde{\bm{H}}_{i}\bm{D}_{i}^{1/2}{\bm{T}_i}^{1/2} {\bm{X}_i} + \bm{W}_l.
\end{equation}
In Sec.\ref{sec4}, we will first focus on the channel model~\eqref{eq: system model 1} and for the general case, i.e., the channel model~\eqref{eq: system model 13}, we will treat it afterwards.

\subsection{Subspace-based channel estimation}
In massive MIMO, by using subspace-based estimation approach, each user's channel can be estimated blindly and accurately up to a scalar ambiguity, specifically, by identifying the signal subspace from interference subspace. This is possible because when the number of receive antennas is very large  and the  uncorrelated channel is assumed, the signals of different users are projected onto quasi-orthogonal subspaces. Another key condition that makes it possible to split the spectrum into signal and interference eigenvalue clusters, is due to the fact that inside-cell users and outside-cell users exhibit difference in the received power at the intended BS (e.g., due to path loss). These observations have been recently employed in massive MIMO~\cite{EVD},~\cite{blind_pilot_cont} to circumvent the problem of pilot contamination.

Applying singular value decomposition (SVD) to the received signal~\eqref{eq: system model 12} (see~\cite{blind_pilot_cont} for more details), the first $K$ eigenvectors, denoted $\bm{U}_s\in \mathcal{C}^{M\times K}$, can then be identified, which correspond to the signal subspace. These eigenvectors represent an estimate of the channel up to scaling ambiguity which can be resolved by exploiting a short training sequence (e.g., 1 pilot) sent by each user. Having identified $\bm{U}_s$, the received signal is then projected onto $\bm{U}_s$, by which most of interference is annihilated and most of the thermal noise is removed. In this process, a new channel model is obtained:
\begin{eqnarray}\label{xxx}
\tilde{\bm{Y}}_l= \bm{U}_s^\dagger \bm{Y}_l = {\bm{G}}_l{\bm{X}_l} + \tilde{\bm{W}}_l
\end{eqnarray}
where ${\bm{G}}_l\in \mathcal{C}^{K \times K}$ and $\tilde{\bm{W}}_l \in \mathcal{C}^{K \times N}$ are the subspace channel and subspace noise matrix, respectively. Note that the original channel $\bm{H}_{l}$ is not needed for the detection of ${\bm{X}_l}$, rather it can be estimated after ${\bm{X}_l}$ being detected for the purpose of downlink precoding, for example.

\section{Performance analysis}\label{sec4}

In this section we will be primarily concerned with deriving the asymptotic eigenvalue distribution (AED) of the channel models~\eqref{eq: system model 13} and~\eqref{eq: system model 1} with the aid of RMT and FPT. In fact, the density of the distribution is not explicitly derived. Instead, a fixed-point equation that satisfies the Stieltjes transform\footnote{Also known as the Cauchy transform, which encodes all the moments of the underlying distribution in a polynomial function. For a distribution function $f(x)$, its Stieltjes transform is defined by $G(s)= \int_{ - \infty }^{ + \infty } \frac{f(x)dx}{x  - s}, \Im{s} \ne 0$.}~\cite{Muller:2006:RMM:2263243.2267030},~\cite{Couillet:2011:RMM:2161679},~\cite{Tulino:2004:RMT:1166373.1166374} of the density is given, which will serve as grounds for the derivation of the formula that helps identify the support of the distribution, in particular, the gap between the desired signal and interference subspaces. In Sec.\ref{SpecialModel} we shall start with the channel model~\eqref{eq: system model 1} and the channel model~\eqref{eq: system model 13} is treated in Sec.\ref{GeneralModel}.

\subsection{Physical channel model with identical AoAs}\label{SpecialModel}
First, we should stress that the steering matrix $\bm{S}$ in~\eqref{eq: system model 1} is random, but fixed, which is dependent on the physical environment only. Equation~\eqref{eq: system model 1} can be written in the compact form as
\begin{eqnarray}\label{eq:information_plus_noise}
\bm{Y}_l =  \bm {S} \tilde{\bm{H}} \tilde{\bm{D}}^{1/2}{\bm{X}} + \bm{W}_l
\end{eqnarray}
where $\tilde{\bm{H}}=(\tilde{\bm{H}}_1, \tilde{\bm{H}}_2, \cdots,\tilde{\bm{H}}_L)\in \mathcal{C}^{P\times KL}$, $\bm{X}= ( \bm{X}_1, \bm{X}_2, \cdots,\bm{X}_L  )^T \in \mathcal{C}^{KL \times N}$, and $\tilde{\bm{D}}\in\mathcal{R}^{KL \times KL}$ is a diagonal matrix, where the components of the row vectors $(p_{1i}\beta_{1i},\cdots,p_{Ki}\beta_{Ki}), i=1,2,\cdots,L$ are placed on its diagonal.

If the true data covariance matrix $\Sigma$ is assumed available at the BS, then the eigenvalue distribution of the channel can be accurately determined, for instance by $\Sigma = (\bm {S} \tilde{\bm{H}}) \tilde{\bm{D}}(\bm {S} \tilde{\bm{H}})^\dagger + \bm{I}_M$,
but in practice, one only has access to the sample covariance matrix $\Sigma_N$ computed from $N$ data samples, which serves as an approximation of $\Sigma$. Thus, our objective is to study the eigenvalue distribution of $\Sigma_N$ in the asymptotic sense, where $\Sigma_N$ is given by the following advanced information-plus-noise model:
\begin{eqnarray}\label{eq:Cov matrix}
\Sigma_N =(\bm {S} \tilde{\bm{H}} \tilde{\bm{D}}^{1/2}{\bm{X}} + \bm{W}_l)(\bm {S} \tilde{\bm{H}} \tilde{\bm{D}}^{1/2}{\bm{X}} + \bm{W}_l)^\dagger/N
\end{eqnarray}
which is difficult to handle by using Stieltjes transform approach~\cite{Dozier:2007},~\cite{Couillet:2011:RMM:2161679}.
\begin{remark}\label{rem1}
The analysis of the eigenvalue distribution of $\Sigma_N$ under an arbitrary diagonal matrix $\tilde{\bm{D}}$ does not admit a tractable solution. Note that our primary concern here is to investigate the impact of power difference between the desired signal and the interference which guarantees, along with other system parameters, the separability of the signal and the interference subspaces. Therefore, without loss of generality, we assume the worst case scenario of power imbalance between inside-cell and outside-cell users. Specifically, we set all interference powers to the maximum interference power, denoted $P_I$, among all interfering users. On the other hand, we set the powers of all desired signals to the minimum signal power, denoted $P_s$, among all inside-cell users. Based on the above assumption, the diagonal matrix $\tilde{\bm{D}}$ is now comprised of two distinct masses, namely $P_s$ and $P_I$, with multiplicity of $K$ and $K(L - 1)$, respectively.
\end{remark}
%
\begin{enumerate}
\item \emph{Assumption}
\end{enumerate}
\smallskip

For the sake of analytical tractability and asymptotic results, we shall make the following assumptions:
\begin{itemize}
\item[--] The input matrix $\bm{X}$ consists of i.i.d. Gaussian entries. This assumption is not strict since the asymptotic result still holds given the entries are independent with finite moments of order greater than 2~\cite[Section 5.1]{Couillet:2011:RMM:2161679}, which is fulfilled for most signal constellations.
\item[--] The diagonal matrix $\tilde{\bm{D}}$ has bounded eigenvalues, and in the large limit, the eigenvalue distribution converges to a deterministic distribution. Note that this is true since the attenuation due to path loss and shadowing is generally constant over many symbol intervals.
\item[--] The entries of matrix $\bm{S}$ are assumed independent. Note that this assumption is violated in practice, where it is shown in~\cite{Ryan08randomvandermonde} that the moments of Vandermonde matrix are bounded above and below by the moments of Mar\v{c}enko-Pasture distribution~\cite{Marcenko:1967} and Poisson distribution, respectively. Nevertheless, when the number of receive antennas grows large, this assumption becomes reasonable~\cite{Muller:2006:RMM:2263243.2267030}, and hence the moments of Mar\v{c}enko-Pasture distribution will become a good approximation of the moments of $\bm{S}$.
\end{itemize}
\smallskip
\goodbreak
\begin{enumerate}[resume]
\item \emph{One-sided spectral analysis}
\end{enumerate}
\smallskip

While the exact solution of the AED of~\eqref{eq:Cov matrix} is difficult to obtain, to simplify the derivation we assume high $\mathrm{SNR}$ regime, i.e., $\bm{W}_l=0$.  Further, we also assume that $K\ll P$ and $P$ is sufficiently large so that orthogonality between users' channel vectors can still hold\footnote{It is shown in~\cite{6375940} that when $M\to\infty$, spatial correlation yields only a minor penalty on the orthogonality condition compared to the independently distributed channel. Note also for two channels $\bm{h}_{ki}$ and $\bm{h}_{lj}$ as defined in~\eqref{eq:steering} with identical steering matrix $\bm{S}$,
${\lim_{P,M \to \infty}} {{\bm{h}_{ki}^\dagger \bm{h}_{lj}}/{M}} \to 0$ almost surely as $K \ll P$.}. Under these assumptions, the eigenvalue distributions of the signal and the interference can be studied separately.

\begin{prop}\label{thm:main theorem}
Let $\bm{S} \in \mathcal{C}^{M \times P}$, $\bm{\tilde H}_{l} \in \mathcal{C}^{P \times K} $, and ${\bm{X}_l}\in \mathcal{C}^{K\times N}$ be defined as in~\eqref{eq:information_plus_noise}. Also let $M$, K, P, $N \to \infty$ with $K \ll P$, ${K}/{M} \to \alpha$, ${P}/{M} \to \beta$, and ${K}/{N} \to \gamma$. Consider the $ N \times N$ Hermitian matrix $\bm{F} {=} {P_s ( \bm{S} \bm{\tilde H}_{l} \bm{X}_l )^\dagger ( \bm{S} \bm{\tilde H}_{l} \bm{X}_l)}/{MN}$. Then the AED of $\bm{F}$ converges to a non-random distribution with Stieltjes transform $G_{\bm{F}}(s)$ satisfying the following equation:
\begin{align}\label{eq:main result}
\beta \gamma^2 (s G_{\bm{F}}(s) &+ 1)  + P_s G_{\bm{F}}(s)(s G_{\bm{F}}(s)-\gamma+1) ( \alpha s G_{\bm{F}}(s) \nonumber \\
 &+ \alpha-\gamma  ) (\alpha s G_{\bm{F}}(s)+\alpha-\beta \gamma) = 0.
\end{align}
\end{prop}

\begin{IEEEproof}
consider the scaled matrix product
\begin{equation}\label{eq:normalizd matrix product}
\bm{D}_k=\sqrt{{a_k}/{m p n}}{\bm{A}_k\bm{B}_k\bm{C}_k}
\end{equation}
and let
\begin{equation}\label{eq:D_k^H D_k}
\bm{G}_1 \stackrel{\Delta}{=}\bm{D}_k^\dagger \bm{D}_k={a_k} \bm{C}_k^\dagger(\bm{A}_k\bm{B}_k)^\dagger (\bm{A}_k\bm{B}_k) \bm{C}_k/{m p n}
\end{equation}
where $a_k$ is a multiplicative constant, and the matrices $\bm{A}_k\in \mathcal{C}^{m\times p}$, $\bm{B}_k \in \mathcal{C}^{p\times l_k}$, and  $\bm{C}_k \in \mathcal{C}^{l_k\times n}$ are mutually independent. Moreover, the entries of $\bm{A}_k$ are assumed independent with zero mean and unit variance, whereas  the entries of $\bm{B}_k$ and $\bm{C}_k$ are assumed i.i.d. $\mathcal{CN}(0,1)$. We also assume $\operatorname{rank}\left ({\bm{A}_k \bm{B}_k \bm{C}_k}\right )= l_k$, i.e., $\bm{G}_1$ has $l_k$ nonzero eigenvalues.

Following~\cite{Muller:2006:RMM:2263243.2267030}, note that $\mathop{tr}({\bm{C}_k^\dagger (\bm{A}_k \bm{B}_k)^\dagger (\bm{A}_k\bm{B}_k) \bm{C}_k})= \mathop{tr}({(\bm{A}_k\bm{B}_k)^\dagger (\bm{A}_k\bm{B}_k) \bm{C}_k}{\bm{C}_k^\dagger})$, i.e., both product factors inside the trace operators share the same nonzero eigenvalues and differ only in $\left| n - l_k \right|$ zeros. Thus it would be more adequate to study the eigenvalue distribution of the $l_k \times l_k$ matrix
\begin{equation}\label{final marix product}
\bm{G}_2\stackrel{\Delta}{=} {a_k} (\bm{A}_k\bm{B}_k)^\dagger (\bm{A}_k\bm{B}_k)(\bm{C}_k \bm{C}_k^\dagger)/m p n
\end{equation}
without the need to obtain the eigenvalue distribution of $\bm{G}_1$ directly. Now let  $\bm{G}_2=\bm{G}_{21}\bm{G}_{22}$, where $\bm{G}_{21}$ and $\bm{G}_{22}$  are defined, respectively, as
\begin{align}
\bm{G}_{21} &\stackrel{\Delta}{=}{a_k} (\bm{A}_k\bm{B}_k)^\dagger (\bm{A}_k\bm{B}_k)/{m p}\label{eq: G21}, \\
\bm{G}_{22} &\stackrel{\Delta}{=} \bm{C}_k \bm{C}_k^\dagger/{n}\label{eq: G22}.
\end{align}

To find the AED of $\bm{G}_2$, one may first derive the marginal distributions of $\bm{G}_{21}$ and $\bm{G}_{22}$ by  means of RMT, then FPT, namely, the free multiplicative law~\cite{Speicher_1991},~\cite{Speicher2006},~\cite{Tulino:2004:RMT:1166373.1166374} can be invoked straightforwardly. The usefulness of FPT lies in the possibility of computing the AED of a product and a sum of Hermitian random matrices based solely on the marginal distributions, given the  matrices are asymptotically free (see e.g.,~\cite[Sec. 2.4]{Tulino:2004:RMT:1166373.1166374} for details on asymptotic freeness of random matrices). In FPT, the free multiplicative law of a product of random matrices is computed via the so-called $S$-transform (see e.g.,~\cite{Tulino:2004:RMT:1166373.1166374}, Def. 2.15.). More specifically, when two Hermitian random matrices are asymptotically free, with each matrix has an eigenvalue distribution that converges almost surely in distribution to a compactly supported distribution, then the $S$-transform of their product is the product of their corresponding $S$-transforms [\cite{Couillet:2011:RMM:2161679}, Thm. 4.7].

Under the assumptions made on $\bm{A}_k$ and $\bm{B}_k$, the asymptotic freeness condition is fulfilled (see Appendix for details). From~\cite{Muller:2006:RMM:2263243.2267030}, the Stieltjes transform $G_{\bm{G}_{21}}(s)$ of the AED of $\bm{G}_{21}$ satisfies the following equation:
\begin{align}\label{eq:Stieltjes transform of G21}
a_k \alpha_1^2 &s^2 G_{\bm{G}_{21}}^3(s) + a_k \alpha_1 s ( \alpha_2 + 1 -2 \alpha_1 ) G_{\bm{G}_{21}}^2(s) \\
&+ ( \alpha_2 s  + a_k(\alpha_1-1 ) (\alpha_1 - \alpha_2) ) G_{\bm{G}_{21}}(s) - \alpha_2 = 0 \nonumber
\end{align}
where $\alpha_1 = {l_k}/{m}$, and $\alpha_2 = {p}/{m}$ are limiting ratios.
Further, from~\cite[Def. (21)]{Muller:2006:RMM:2263243.2267030}\footnote{Note that different definition of Stieltjes transform is used in~\cite{Muller:2006:RMM:2263243.2267030}.}, if we substitute $s = -1/\lambda$ in~\eqref{eq:Stieltjes transform of G21}, and replace $G_{\bm{G}_{21}}(-1 / \lambda)$ by $-\lambda (\Upsilon(\lambda)+1)$, we obtain
\begin{align}\nonumber
&a_k\alpha_1^2\lambda (\Upsilon ( \lambda ) + 1)^3  +  a_k \alpha_1 \lambda ( \Upsilon ( \lambda ) + 1 )^2 ( \alpha_2 - 2\alpha_1  + 1 )\\
& - \lambda ( {\alpha_2}/{\lambda} - a_k ( \alpha_1 - \alpha_2 ) (\alpha_1 - 1 ) )  (\Upsilon ( \lambda) + 1 )  +  \alpha_2=0.
 \end{align}
Replacing $\lambda$ by $\Upsilon^{-1}(z)$, and using~\cite[Def. (20)]{Muller:2006:RMM:2263243.2267030}, it is easy to check that the $S$-transform $S_{\bm{G}_{21}}(z)$ reads
\begin{eqnarray}\label{eq:S trnasform of G21}
S_{\bm{G}_{21}} (z) = \frac{\alpha_2}{ a_k( \alpha_1 z + \alpha_2) (\alpha_1 z + 1)}.
\end{eqnarray}
\par
Next, since the entries of $\bm{C}_{k}/\sqrt{n}$ are assumed i.i.d. each with zero mean and variance ${1}/{n}$, then $\bm{G}_{22}$ is a central Wishart matrix with $n$ degrees of freedom~\cite{Hiai:2006:SLF:1204180},~\cite{Couillet:2011:RMM:2161679},~\cite{Tulino:2004:RMT:1166373.1166374}. Further, its eigenvalue distribution follows the well-known Mar\v{c}enko-Pastur law~\cite{Marcenko:1967}. Thus when $ n, l_k \to \infty $ with a fixed ratio ${l_k}/{n} \to \alpha_3$, the $S$-transform $S_{\bm{G}_{22}} (z)$ corresponding to $\bm{G}_{22}$ reads as~\cite{Couillet:2011:RMM:2161679},~\cite{Tulino:2004:RMT:1166373.1166374}
\begin{eqnarray}\label{eq: S transform of G22}
S_{\bm{G}_{22}} (z) =\frac{1}{(\alpha_3 z + 1)}.
\end{eqnarray}

Returning to~\eqref{final marix product}, it is easy to see that, as a consequence of asymptotic freeness of $\bm{G}_{21}$ and $\bm{G}_{22}$ (the validity of this assumption is discussed in details in Appendix), the $S$-transform corresponding to ${\bm{G}_{2}}$ is the product of $S_{\bm{G}_{21}} (z)$ and $S_{\bm{G}_{22}} (z)$, that is~\cite[Thm. 4.7]{Couillet:2011:RMM:2161679}
\begin{eqnarray}\label{eq: S transform of G21G22}
S_{\bm{G}_{2}} (z) = \frac{\alpha_2} { a_k(\alpha_1 z + \alpha_2 )(\alpha_3 z + 1)(\alpha_1 z + 1)},
\end{eqnarray}
and from~\cite[Thm. 2.32]{Tulino:2004:RMT:1166373.1166374}, it can be shown by straightforward algebra the $S$-transform corresponding to $\bm{G}_1$~\eqref{eq:D_k^H D_k} is
\begin{equation} \label{eq:S-transform of D^H*D}
S_{\bm{G}_1} (z) = \frac{\alpha_2}{a_k (\frac{\alpha_1}{\alpha_3} z+\alpha_2   )(\frac{\alpha_1}{\alpha_3} z + 1)( z+\alpha_3) }.
\end{equation}
\par
Now we are left with the task of determining the Stieltjes transform corresponding to~\eqref{eq:S-transform of D^H*D}. From~\cite[Def. (20) and (21)]{Muller:2006:RMM:2263243.2267030}, it can be easily verified that the $S$-transform $S_F(z)$ and the Stieltjes transform $G_F(s)$ of a distribution function $F$ satisfies
\begin{eqnarray}\label{eq:S_transform and Stieltjes transform link}
S_F \left(-s G_F(s)-1\right)  =\frac{G_F(s)}{s G_F(s)+1}.
\end{eqnarray}
Using~\eqref{eq:S-transform of D^H*D} and with the aid of~\eqref{eq:S_transform and Stieltjes transform link}, it follows straightforwardly after some mathematical manipulations that the Stieltjes transform $G_{\bm{G}_1}(s)$ corresponding to $\bm{G}_1$ satisfies the following equation:
\begin{align}\label{eq:NNST} \nonumber
\alpha_2 \alpha_3^2 (1 &+ s G_{\bm{G}_1} ) + a_k G_{\bm{G}_1} (1 - \alpha_3 + s G_{\bm{G}_1})
( \alpha_1 \\
&- \alpha_3 + \alpha_1 s G_{\bm{G}_1}) (\alpha_1 - \alpha_2 \alpha_3 + \alpha_1 s G_{\bm{G}_1} )=0
\end{align}

Finally, we remark that if we replace $a_k$, $\alpha_1$, $\alpha_2$ and $\alpha_3$ in~\eqref{eq:NNST} by $P_s$, $\alpha$, $\beta$, and $\gamma$, respectively, then we obtain~\eqref{eq:main result}.
\end{IEEEproof}

Similarly, following the preceding derivation it can be shown that the eigenvalue distribution of interference satisfies~\eqref{eq:NNST} in the Stieltjes transform with $a_k$, $\alpha_1$, $\alpha_2$ and $\alpha_3$ are replaced by $P_I$, ${K(L-1)}/{M}$, $P/M$, and ${K(L-1)}/{N}$, respectively. When the i.i.d. assumption on the channel $\bm{S}\bm{\tilde H}_{l}$ holds we have
\begin{cor}\label{collary0000}
Let $M$, $K$, and $N$ are large with $\alpha$, and $\gamma$ fixed yet not very small. Under rich scattering propagation (i.e., $P\to \infty)$, the eigenvalue distribution of
$\bm{F} = {P_s} (\bm{S}\bm{\tilde H}_{l}{\bm{X}_l})^\dagger (\bm{S}\bm{\tilde H}_{l}{\bm{X}_l})/{MN}$ converges to a non-random distribution with Stieltjes transform $G_{\bm{F}} $ satisfying
\begin{equation}\label{eq:cor1}
\gamma (1 +s G_{\bm{F}}) - P_s G_{\bm{F}} (1 + s G_{\bm{F}} - \gamma ) (\alpha + \alpha s G_{\bm{F}} - \gamma )   = 0.
\end{equation}
\end{cor}
\begin{remark}
 Equation~\eqref{eq:cor1} exactly reproduces the result in~\cite [Eq.(75)]{blind_pilot_cont}, where it was derived under uncorrelated channel model and the convention that $P_s=N/K$.
\end{remark}
\begin{cor}\label{collary1111}
Assume the channel model~\eqref{eq: system model 1}. Further, assume $K, P$ and $N$ are large with $K/P \to \alpha^\prime$ and $\gamma$ fixed but not very small. Then adding more antennas at the BS, the eigenvalue distribution of $\bm{F} = P_s (\bm{S}\bm{\tilde H}_{l}{\bm{X}_l})^\dagger (\bm{S}\bm{\tilde H}_{l}{\bm{X}_l})/{MN}$ converges to a non-random distribution resembling the eigenvalue distribution under uncorrelated channel with $P$ receive antennas. In Stieltjes domain, the AED of $\bm{F}$ satisfies~\eqref{eq:cor1} with $\alpha$ replaced by $\alpha^\prime$.
\end{cor}
\begin{remark}
From Corollaries~\ref{collary0000} and~\ref{collary1111} we observe that the gap between signal subspace and interference subspace under a physical channel with $P$ AoAs and unlimited number of receive antennas $M$ is equivalent to the gap under uncorrelated channel with $M=P$ antennas at the BS. Since blind subspace method relies entirely on the distance between the two subspaces, it turns out that its performance depends mainly on $P$ rather than $M$. Nevertheless, the array gain due to the increase of $M$ can still be partly retained, especially when signal subspace is detached from interference subspace. After that, the performance will be limited by the degrees of freedom of the channel, and hence it saturates.
\end{remark}
\goodbreak
\smallskip
\begin{enumerate}[resume]
\item \emph{The support of the distribution}
\end{enumerate}
\smallskip

In this part we will characterize the support of the distribution, which is important for knowing under what values of the system's parameters the distribution yields two distinct eigenvalues clusters, and hence subspace separability. It is known that the Stieltjes transform is increasing on intervals on the real line outside the support of its distribution function~\cite{Couillet:2011:RMM:2161679}. Further, its inverse function $s(x)$\footnote{The inverse function $s(x)$ is obtained as a solution for variable $s$ in the fixed-point equation satisfying the Stieltjes transform, where $x$ is a real dummy variable.} is also increasing in these intervals only. Therefore, the endpoints of the support can be determined by finding the local extrema of the inverse function $s(x)$. Unfortunately, these local extrema are notoriously difficult to express in a closed-form solution. So this paper is only devoted to deriving approximate formula without explicitly obtaining the extreme points. To that end, we plot $s(x)$ for real $x$ to find the regions where $s(x)$ in increasing, and hence the support of the distribution, denoted $\mathcal{S}$ can be defined as the finite union of regions on the real line where $s(x)$ is not increasing, that is
\begin{equation} \label{eq:supportdefinition}
\mathcal{S}=\mathcal{R}\backslash \bigcup \limits_{{x_1}\l{x_2} \in \mathcal{R} \hfill\atop
\scriptstyle{x_1} < {x_2}\hfill}\{s(x_1),s(x_2)|\forall x \in (x_1,x_2),\frac{d}{dx}s(x)> 0 \}.
\end{equation}

To find the support of the distribution, it seems more adequate to use the $l_k\times l_k$ matrix $\bm{G}_2$ (i.e., a matrix of much smaller dimensions) rather than the $n \times n$ matrix $\bm{G}_1$, since we are only interested in the nonzero eigenvalues. By the aid of~\eqref{eq:S_transform and Stieltjes transform link} one can verify that the Stieltjes transform $G_{\bm{G}_2}$  corresponding to the $S$-transform~\eqref{eq: S transform of G21G22} satisfies
\begin{align}\label{eq:StieltjesTransformOfKK}  \nonumber
s G_{\bm{G}_2} + a_k G_{\bm{G}_2} (\frac{l_k}{m} s G_{\bm{G}_2} &+ \frac{l_k}{m} - 1 ) (\frac{l_k}{n} s G_{\bm{G}_2} + \frac{l_k}{n} -1)\times \\
&(\frac{l_k}{p} s G_{\bm{G}_2} + \frac{l_k}{n} -1)+ 1 =0.
\end{align}
By expanding~\eqref{eq:StieltjesTransformOfKK} and after simple mathematical manipulations we have
\begin{dmath}\label{eq:invfunction1}
a_k  l_k^3 x^4 s^3 - a_k  l_k^2 x^3 \left(-3 l_k + M + N + P\right) s^2
+x \left(M N P + a_k l_k x \left (3 l_k^2 + M N + \left(M + N\right) P - 2 l_k \left(M + N + P\right)\right) \right)s
+M N P - a_k x \left(-l_k + M \right) \left(l_k - N \right) \left(l_k - P\right)=0
\end{dmath}
where we have replaced $G_{\bm{G}_2}$ by the dummy variable $x$ and $m$, $n$, and $p$ by $M$, $N$, and $P$, respectively. Note that~\eqref{eq:invfunction1} is  a cubic equation in $s$, and its three roots, denoted $s(x)=\{s_1(x), s_2(x), s_3(x)\}$ must be computed, and hence the support is defined by~\eqref{eq:supportdefinition}. We remark that~\eqref{eq:invfunction1} can be also used to find the support of eigenvalue distribution of interference by noting that $a_k=P_I$ and  $l_k=K(L-1)$.
\goodbreak
\smallskip
\begin{enumerate}[resume]
\item \emph{Double-sided spectral analysis}
\end{enumerate}
\smallskip

So far we have treated the signal subspace and the interference subspace separately. However, as far as systems with finite dimensions are concerned in practice, the two sets of eigenvectors corresponding to the signal and the interference are somehow interconnected. Thus the eigenvalue distribution of the signal and the interference, should be studied jointly. For the sake of mathematical tractability, we again assume high $\mathrm{SNR}$ regime, i.e., we set $\bm{W}_l=0$ in~\eqref{eq:Cov matrix}. Let
\begin{equation}\label{eq:G1}
\tilde {\bm{G}}_1\stackrel{\Delta}{=}  (\bm {S} \tilde{\bm{H}} ) \tilde{\bm{D}}^{1/2}{\bm{X}}{\bm{X}}^\dagger \tilde{\bm{D}}^{1/2}(\bm {S} \tilde{\bm{H}} )^\dagger /MN.
\end{equation}
Instead, it is more convenient to study the eigenvalue distribution of
\begin{equation}\label{eq:SJG2prime}
\tilde{\bm{G}}_2 \stackrel{\Delta}{=}  \tilde{\bm{D}}^{1/2}{\bm{X}}{\bm{X}}^\dagger \tilde{\bm{D}}^{1/2} (\bm {S} \tilde{\bm{H}} )^\dagger  (\bm {S} \tilde{\bm{H}} )/MN
\end{equation}
without accessing to the information about the eigenvalues of $\tilde {\bm{G}}_1$. Further, let $\tilde{\bm{G}}_2=\tilde {{\bm{G}}}_{21}\tilde {{\bm{G}}}_{22}$, where $\tilde {{\bm{G}}}_{21}$ and $\tilde {{\bm{G}}}_{22}$ are defined as follows:
\begin{align}
\tilde {{\bm{G}}}_{21}&\stackrel{\Delta}{=}\tilde{\bm{D}}^{1/2}{\bm{X}}{\bm{X}}^\dagger \tilde{\bm{D}}^{1/2} /N,\label{AAAA} \\
\tilde {{\bm{G}}}_{22}&\stackrel{\Delta}{=} (\bm {S} \tilde{\bm{H}} )^\dagger  (\bm {S} \tilde{\bm{H}} )/M \label{BBB}
\end{align}

\begin{remark}
Equation~\eqref{eq:SJG2prime} is indeed intuitive in the sense that it gives an insight into the behavior of the gap between the signal subspace and the interference subspace as the ratio $P_s/P_I$ varies. To see this, assume a rich scattering environment, i.e., $P\to \infty$. From central limit theorem, as $N,M \to \infty$ with $K, L$ fixed, $\bm{X}{\bm{X}}^\dagger /N$, and  ${(\bm {S} \tilde{\bm{H}})^\dagger (\bm {S} \tilde{\bm{H}})}/M $ reduce to identity matrices. Thus the final spectrum of~\eqref{eq:SJG2prime} is dictated by the spectrum of $\tilde{\bm{D}}$, that is, a probability mass function (pmf) with two masses at $P_s$ and $P_I$. In this case, subspace separability is possible whenever $P_s>P_I$. However, when the dimensions of random matrices in~\eqref{eq:SJG2prime} grow large, but with fixed ratios the eigenvalue distribution converges to a non-random distribution rather than the pmf. This leads to eigenvalues spread around the center eigenvalues\footnote{The center eigenvalues are determined by the deterministic quantities in the channel model. For instance, ${Y_l Y_l^\dagger}/{MN}$ has two deterministic quantities, namely, $P_s$ and $P_I$ corresponding to the effective signal and interference powers, respectively.}. Therefore, a large difference between those deterministic parameters of the system is important which help to shift the two clusters of eigenvalues away from each other.
\end{remark}
\par
From~\cite{Silverstein1995175}, the asymptotic Stieltjes transform of ${\bm{X}^\dagger \tilde{\bm{D}}{\bm{X}}}/N$, denoted $G_{\bm{X}^\dagger \tilde{\bm{D}}{\bm{X}}/N}(s)$, as $KL$, $N \to \infty$ with ${KL}/{N}\to \gamma$ is the unique solution of
\begin{eqnarray}\label{eq:integral}
G_{\bm{{\bm{X}}^\dagger \tilde{\bm{D}}{\bm{X}}}/N}(s)=-\left ( s-\gamma \int \frac{\lambda f(\lambda) d \lambda}{1+\lambda G_{\bm{{\bm{X}}^\dagger \tilde{\bm{D}}{\bm{X}}}/N}(s)} \right )^{-1}
\end{eqnarray}
where $f(\lambda)$ is the probability density function (pdf) of eigenvalue of $\bm{X}^\dagger \tilde{\bm{D}} \bm{X}/N$. It would seem difficult to evaluate~\eqref{eq:integral}. Since the entries of $\bm{X}$ are assumed i.i.d. $\mathcal{CN}(0,1)$, this class of matrices is unitarily invariant and thus asymptotically free w.r.t. any Hermitian matrix (see Appendix). Then it follows from~\cite [Eq.(43)]{1013149} that
\begin{equation} \label{eq:S_transform of inf plus noise1}
S_{\tilde {{\bm{G}}}_{21}}(z)=\frac{1}{\gamma z + 1}S_{\tilde{\bm{D}}}(z)
\end{equation}
which is derived within the framework of FPT.
Since the eigenvalue pdf $f_{\tilde{\bm{D}}}(x)$ of $\tilde{\bm{D}}$ converges to a pmf of two distinct eigenvalues, namely, $P_s$ and $P_I$, hence $f_{\tilde{\bm{D}}}(x)$ can be written as
\begin{equation}
f_{\tilde{\bm{D}}}(x)=\frac{1}{L} \delta(x-P_s) + \frac{L-1}{L}\delta(x-P_I)
\end{equation}
where $\delta(.)$ is the Dirac delta function. By the definition of Stieltjes transform~\cite{Muller:2006:RMM:2263243.2267030},~\cite{Couillet:2011:RMM:2161679},~\cite{Tulino:2004:RMT:1166373.1166374}, one can verify that $f_{\tilde{\bm{D}}}(x)$ admits a Stieltjes transform $G_{\tilde{\bm{D}}}(s)$ of the form
\begin{eqnarray}
  G_{\tilde{\bm{D}}}  = \frac{L P_s - L s + P_I - P_s }{L (P_s - s) (P_I - s)}
\end{eqnarray}
and from~\cite [Def. 3.4]{Couillet:2011:RMM:2161679}, it is easy to check that the corresponding $S$-transform fulfils the following quadratic equation
\begin{dmath}\label{S-transform quadratic equation}
 L  P_I  P_s  z  S_{\tilde{\bm{D}}}^2(z) - (P_s - P_I + L  P_I + L  P_I  z + L  P_s  z)  S_{\tilde{\bm{D}}}(z) + ( L + L  z)=0
\end{dmath}
with the two roots
\begin{equation} \label{eq:StransformofD}
S_{\tilde{\bm{D}}}(z) = \frac{b(z)\pm \sqrt{b^2(z)-4 L^2 P_I P_s (z +1)z}}{2 L  P_I  P_s  z}
\end{equation}
where $b(z) \stackrel{\Delta}{=} P_s - P_I + L  P_I + L  P_I  z + L  P_s  z$. One may prove that the root with minus sign is the true solution to~\eqref{S-transform quadratic equation}\footnote{Since the $S$-transform of a distribution with a single mass $P_s$ is ${1}/{P_s}$, when $P_I=P_s$, it follows that the root with minus sign yields $S$-transform of the form ${1}/{P_s}$.}. Let ${KL}/{M} \to \alpha $ and ${KL}/{P} \to \eta$  as $KL$, $M$, and $P\to \infty$. Plugging~\eqref{eq:StransformofD} into ~\eqref{eq:S_transform of inf plus noise1}, combining the result with~\eqref{eq:S trnasform of G21}, and assuming the factors~\eqref{AAAA} and~\eqref{BBB} are asymptotically free (see Appendix), then it follows from~\cite [Thm. 2.32]{Tulino:2004:RMT:1166373.1166374} the $S$-transform corresponding to $\tilde{\bm{G}}_2$  can be finally written as
\begin{equation} \label{eq:5}
S_{\tilde{\bm{G}}_2}(z) = \frac{b(z)- \sqrt{b^2(z)-4 L^2 P_I P_s (z +1)z}}{2 L P_I P_s z (\gamma z + 1)(\alpha z + 1)( \eta z + 1)}.
\end{equation}

By making use of~\eqref{eq:S_transform and Stieltjes transform link} in~\eqref{eq:5} we have the following result.
\begin{prop}\label{prop:prop2}
Let $\bm{S}$, $\tilde{\bm{H}}$, $\tilde{\bm{D}}$, and $\bm{X}$ be defined as in~\eqref{eq:information_plus_noise}. Let also $KL$, $N$, $P$, and $M \to \infty$, with ${KL}/{M}\to \alpha$, ${KL}/{P}  \to \eta$, and ${KL}/{N} \to \gamma$. Further, denote $G_{\tilde{\bm{G}}_2}$ the Stieltjes transform corresponding to $\tilde{\bm{G}}_2$~\eqref{eq:SJG2prime}. Then $G_{\tilde{\bm{G}}_2}$ is the unique solution of the following fixed-point equation:
\begin{align}\label{eq:fullSJ}\nonumber
LP_s &(s G_{\tilde{\bm{G}}_2} +1) - P_s + P_I (1 + L (s G_{\tilde{\bm{G}}_2} \nonumber
 + 2 G P_s ( \alpha s G_{\tilde{\bm{G}}_2} \\\nonumber
 &+ \alpha -1 )
(\eta s G_{\tilde{\bm{G}}_2} + \eta -1 ) (\gamma s G_{\tilde{\bm{G}}_2} + \gamma -1))) \\ \nonumber
 &+ ( P_I^2 ( L s G_{\tilde{\bm{G}}_2} +1  )^2  + P_s^2 ( L s G_{\tilde{\bm{G}}_2} + L - 1  )^2\\
 &- 2 P_I P_s (1 - L + L^2( s G_{\tilde{\bm{G}}_2} + 1 ) s G_{\tilde{\bm{G}}_2}  )  )^{1/2} = 0.
\end{align}
\end{prop}
Using~\eqref{eq:fullSJ} to find the support of the distribution is difficult, hence, we adopt an approximate solution. Rewriting~\eqref{eq:fullSJ} in terms of the parameters $K$, $L$, $M$, $N$, and $P$ yields
\begin{align}\label{eq:roots_of_Q}\nonumber
&\frac{2 K^3 L^4 P_I P_s}{M N P} x \upsilon^3 - 2  K^2 L^3  {P_I}  {P_s} x (\frac{1}{M N} +\frac{1}{M P} +\frac{1}{N P} )\upsilon^2 \\ \nonumber
&+({2  K L^2  {P_I}  {P_s}}x (\frac{1}{M}  + \frac{1}{N}  +\frac{1}{P})  + L  ({P_I}+   {P_s}))\upsilon \\ \nonumber
&+ (P_I^2 (L (\upsilon-1)+1)^2 + P_s^2 (L \upsilon - 1)^2 \\ \nonumber
&+ 2 P_I P_s (L - L^2 ( \upsilon - 1) \upsilon - 1) )^{1/2} \\
&-L  {P_I}+ {P_I}- {P_s}- 2  L  {P_I}  {P_s} x=0
\end{align}
where we have replaced $G_{\tilde{\bm{G}}_2}$ by the dummy variable $x$ and $\upsilon=s x+1$. Note that~\eqref{eq:roots_of_Q} (from which the all roots {$s(x)$} must be found and all extreme values are identified) is a polynomial of degree 6 in variables $s$ and $x$, hence, finding its zeros is indeed intractable. Instead, we adopt an approximate solution such that the above higher order polynomial is reduced to a low-order polynomial of degree 2 in variable $s$. To that end, notice that the multiplicative coefficients of the terms $\upsilon^3$, $\upsilon^2$ and $\upsilon$ scale like $\alpha \eta \gamma$, $\alpha \gamma +\alpha \eta +\eta \gamma$, and $\alpha+\eta+\gamma$, respectively. It turns out that in order to truncate out the terms $\upsilon^3$ and $\upsilon^2$ and hence have a good approximation of the true solution, the following conditions should be satisfied:
\begin{eqnarray}
\frac{\alpha+\eta+\gamma}{\alpha \eta \gamma} \gg1, \frac{\alpha+\eta+\gamma}{\alpha \gamma +\alpha \eta +\eta \gamma} \gg 1
\end{eqnarray}
where ignoring the terms $\upsilon^3$ and $\upsilon^2$ don't affect the final answer significantly. Therefore, after truncating the higher order terms, namely, $\upsilon^3$ and $\upsilon^2$, we have
\begin{align}\label{eq:invfunction2} \nonumber
({2  K L^2  {P_I}  {P_s}}&x(\frac{1}{M}  + \frac{1}{N}  +\frac{1}{P})  + L  \left({P_I}+   {P_s}\right))\upsilon \\ \nonumber
&+ (P_I^2 (1 + L (\upsilon-1))^2 + P_s^2 (L \upsilon - 1)^2 \\ \nonumber
&+2 P_I P_s (L - L^2 (\upsilon-1) \upsilon - 1) )^{1/2}\\
&-L  {P_I}+ {P_I}- {P_s}- 2  L  {P_I}  {P_s} x = 0
\end{align}
and its two roots $s_1(x)$ and $s_2(x)$ can be computed, and thus the support is defined by~\eqref{eq:supportdefinition}.

\subsection{Physical channel model with distinct AoAs}\label{GeneralModel}

 So far we have considered the case of identical AoAs. However, depending on the complexity of the propagation environment, we may have different signals received from different/shared directions, captured by the channel model~\eqref{eq: system model 12}. Indeed, the analytical analysis of this channel model is extremely intricate. For the sake of simplicity, there is no significant loss of generality incurred by using the channel model~\eqref{eq: system model 13}. Here we recall that the received signals of the $i$th cell impinge upon the array of the intended BS from i.i.d. AoAs of cardinality $P_i$. We also assume that all sets of AoAs over all cells are mutually independent. Without loss of generality, we consider the worst case of power imbalance between the desired signal and the interfering signal as has been stated in \emph{Remark}\ref{rem1}. Now we may rewrite~\eqref{eq: system model 13} as
\begin{equation} \label{eq:physicalchannel_2}
\bm{Y}_l=\sqrt{P_s} \bm{S}_l\tilde{\bm{H}}_l\bm{X}_l + \sqrt{P_I}\sum\limits_{i = 1,i\ne l}^L \bm{S}_i\tilde{\bm{H}}_i\bm{X}_i +\bm{W}_l.
\end{equation}

Studying the eigenvalue distribution of the signal model~\eqref{eq:physicalchannel_2} is intractable even when $\bm{W}_l=0$. Therefore, a seemingly natural way is to do one-sided analysis of the distribution of the first and second terms in~\eqref{eq:physicalchannel_2}. This, however, would help us get some insights into the behavior of the eigenvalue distribution under the channel model~\eqref{eq:physicalchannel_2} as it will be shown shortly. For the sake of illustration, we let $l=1$.

First, note that the support of the eigenvalue distribution of the first term in~\eqref{eq:physicalchannel_2} can be computed by~\eqref{eq:invfunction1}. On the other hand, the second term of~\eqref{eq:physicalchannel_2} can be rewritten in a compact way:
\begin{equation}
\bm{Y}_I = \sqrt{P_I} \bm{S}_I\bm{H}_I\bm{X}_I
\end{equation}
where
$\bm{S}_I = (\bm{S}_2, \bm{S}_3, \cdots, \bm{S}_L )\in\mathcal{C}^{M \times n}$
is the composite steering matrix, where $n=P_2+P_3+\cdots+P_L$.
$\bm{H}_I \in\mathcal{C}^{n \times K(L-1)}$ is a block diagonal matrix whose blocks are $\tilde{\bm{H}}_2, \tilde{\bm{H}}_3, \cdots,\tilde{\bm{H}}_L$ and $\bm{X}_I = (\bm{X}_2, \cdots, \bm{X}_L)^T \in\mathcal{C}^{K(L-1) \times N}$ comprises all interference signals. Note that the eigenvalues of $\bm{H}_I$ equals the combined eigenvalues of the submatrices on its diagonal. Moreover, the eigenvalue distribution of each Hermitian matrix $\tilde{\bm{H}}_i\tilde{\bm{H}}_i^\dagger$ converges to a non-random distribution as $K, P_i \to \infty $ with a fixed ratio $K / P_i$. This implies that the eigenvalue distribution of $\bm{H}_I\bm{H}_I^\dagger$ also converges to a non-random distribution. To proceed with the analysis, the following Lemma enables us to characterize the behavior of eigenvalue distribution asymptotically.

\begin{lem}\label{lemmmm}
Assume $K < P_i, i= 2, \cdots,L$ with fixed ratios ${K}/{P_i}\to \beta_i$. Further, let $\lambda_i=P_i/n$ and $f_i(x)$ be the eigenvalue distribution of ${\tilde{\bm{H}}_i \tilde{\bm{H}}_i^\dagger}$. Then, as $K, P_i\to \infty$, the eigenvalue distribution $f(x)$ of $\bm{H}_I \bm{H}_I^\dagger$ is related to $f_i(x), i=2,3,\cdots,L$ through the following relationship:
\begin{equation} \label{lemmaa}
 f(x) =  \sum\limits_{i = 2}^L \lambda_i f_i(x).
 \end{equation}
\end{lem}

\begin{IEEEproof}
Since the eigenvalues of $\bm{H}_I$ are equal to the combined eigenvalues of $\tilde{\bm{H}}_2,\cdots,\tilde{\bm{H}}_L $, we can write
\[ \mathop{tr}(\bm{H}_I\bm{H}_I^\dagger)= \mathop{tr}(\tilde{\bm{H}}_2\tilde{\bm{H}}_2^\dagger)+\cdots+\mathop{tr}(\tilde{\bm{H}}_L\tilde{\bm{H}}_L^\dagger)\]
and the $k$th moment of the eigenvalue distribution of ${\bm{H}_I \bm{H}_I^\dagger}$ is given by
\begin{dmath}\label{trace_moment}
 \mathop{tr}  (\bm{\bm{H}}_I\bm{H}_I^\dagger )^k= \mathop{tr}(\tilde{\bm{H}}_2\tilde{\bm{H}}_2^\dagger)^k + \cdots+\mathop{tr}(\tilde{\bm{H}}_L\tilde{\bm{H}}_L^\dagger)^k.
 \end{dmath}

Note that the number of eigenvalues of $\bm{H}_I\bm{H}_I^\dagger$ are $n$, while each submatrix $\tilde{\bm{H}}_i \tilde{\bm{H}}_i^\dagger$ has $P_i$ eigenvalues (i.e., the assumption $K<P_i$ implies $K$ nonzero eigenvalues and $P_i-K $ zero eigenvalues). Thus the normalized trace form of~\eqref{trace_moment} reads

\begin{dmath}\label{norm_trace}
{\mathop{tr}}_n  (\bm{H}_I\bm{H}_I^\dagger )^k= \frac{1}{n}(P_2 {{\mathop{tr}}_{P_2}(\tilde{\bm{H}}_2\tilde{\bm{H}}_2^\dagger)^k}+\cdots + P_L {{\mathop{tr}}_{P_L} (\tilde{\bm{H}}_L\tilde{\bm{H}}_L^\dagger)^k}).
\end{dmath}
Recall that the Stieltjes transform $G_{\bm{E}}(s)$ of $n \times n$ Hermitian matrix $\bm{E}$ can be expanded in a Laurent series involving the moments of $\bm{E}$ as (see e.g.,~\cite[Thm. 3.3]{Couillet:2011:RMM:2161679})
\begin{equation}\label{laurentseries}
G_{\bm{E}}(s)=-\frac{1}{s} \sum\limits_{k = 0}^\infty \frac{{\mathop{tr}}_{n}(\bm{E}^k)}{s^k}.
\end{equation}
Combining~\eqref{laurentseries} with~\eqref{norm_trace} yields
\begin{equation}\label{lammaaa}
 G_{\bm{H}_I \bm{H}_I^\dagger}(s)= \sum\limits_{i = 2}^L \lambda_i G_{\tilde{\bm{H}}_i \tilde{\bm{H}}_i^\dagger}(s).
\end{equation}
Then the eigenvalue distribution is reconstructed from $G_{\bm{H}_I \bm{H}_I^\dagger}(s)$ by applying the inversion formula of Stieltjes transform~\cite{Tulino:2004:RMT:1166373.1166374},~\cite{Couillet:2011:RMM:2161679} to obtain~\eqref{lemmaa}.
\end{IEEEproof}
\par
Since each submatrix $H_i$ consist of i.i.d. complex Gaussian entries, it follows from~\eqref{lemmaa} that the eigenvalue distribution of $\bm{H}_I\bm{H}_I^\dagger$ is a weighted sum of the Mar\v{c}enko-Pasture laws~\cite{Marcenko:1967} with different ratios $\beta_i, i=2,\cdots,L$. One particular case is when $P_2=\cdots = P_L=P$, it follows that the eigenvalue distribution of $\bm{H}_I\bm{H}_I^\dagger$ reduces to a Mar\v{c}enko-Pasture distribution with a single parameter $\beta = {K}/{P}$.
\begin{prop}\label{finalprop}
Consider the channel model~\eqref{eq:physicalchannel_2}. Without loss of generality, assume $P_1>P_2>\cdots>P_L$. Then the users in the $L$th cell contribute most to the spreading of the eigenvalues of channel. Further,
under any system parameters, the power ratio $P_s/P_I$ should be adjusted according to the $L$th cell, which is required to maintain specified link performance when subspace method is used at BS.
\end{prop}

\begin{remark}
We observe that applying the free multiplicative law to the channel model~\eqref{eq:physicalchannel_2} is extremely difficult because of nonidentical dimensions of the matrices $H_i, i=1,2,\cdots,L$. Therefore, we provide here a non-rigorous yet intuitive proof of Proposition~\ref{finalprop}. Further, the fact that the users experiencing less channel scatters render the eigenvalue distribution getting wider is indeed very intuitive.
Notice that the maximum gap between the two subspaces occurs when $P\to \infty$ (i.e., uncorrelated channel), i.e., ${K}/{P} \to 0$. In addition, from Lemma~\ref{lemmmm}, the eigenvalue distribution of $\bm{H}_I\bm{H}_I^\dagger$ is a weighted sum of Mar\v{c}enko-Pasture distributions with different ratios. Assume that all limiting ratios, $K/P_i$, are infinitely small except one ratio, say the last ratio, which is made large (i.e., the signals of the $L$th cell are received at the desired BS from a very small number of AoAs). Thus the support of the resulting distribution is dominated by the $L$th distribution. This leads to a wider spread of the eigenvalues of the interference.
\end{remark}

To get insight into how the case of distinct AoAs may shape the distribution, and thus change the support of interference eigenvalue distribution, we assume the same number of AoAs per each cell w.r.t. the desired BS. Again by using one-sided spectral analysis and by following similar steps in Sec.\ref{SpecialModel} with the aid of Lemma~\ref{lemmmm}, one can prove that the eigenvalue distribution of the $N\times N$ matrix $\bm{Y}_I^\dagger \bm{Y}_I/MN$  reads in the Stieltjes domain as
\begin{align}\label{xxxxxxxxx} \nonumber
&- N P_I ( 1 +  s x ) ( K - K L + N +  N s x  ) ( N + P - L P \\ \nonumber
&+  N s x ) x + M ( K P P_I x(L-1)^2 - N(  L-1)\times\\
&( P + P_I x  ( K + P )  )  (1 + sx  ) + P_I x ( N +  N s x)^2 )=0.
\end{align}
By computing the three roots of~\eqref{xxxxxxxxx}, $s(x)=\{s_1(x),s_2(x),s_3(x)\}$, then the support is defined by~\eqref{eq:supportdefinition}.
\goodbreak

\section{Numerical results}\label{sec5}
In this section, we use finite-size scenarios to show some numerical results that verify the theoretical results obtained in the asymptotic limit. We also provide simulation results for the uncoded bit error rate (BER) and compare the performance of subspace-based channel estimation scheme~\cite{blind_pilot_cont} under the physical channel model and the i.d. channel. The pilot-based channel estimation scheme is also shown. In our simulations, the BS antennas are sparsely spaced at twice the carrier wavelength unless explicitly stated otherwise.
\subsection{The support of AED}
\begin{figure*}[!ht]
\centering
\subfloat[approximate solution using~\eqref{eq:invfunction1}]{\includegraphics[width=2.7in]{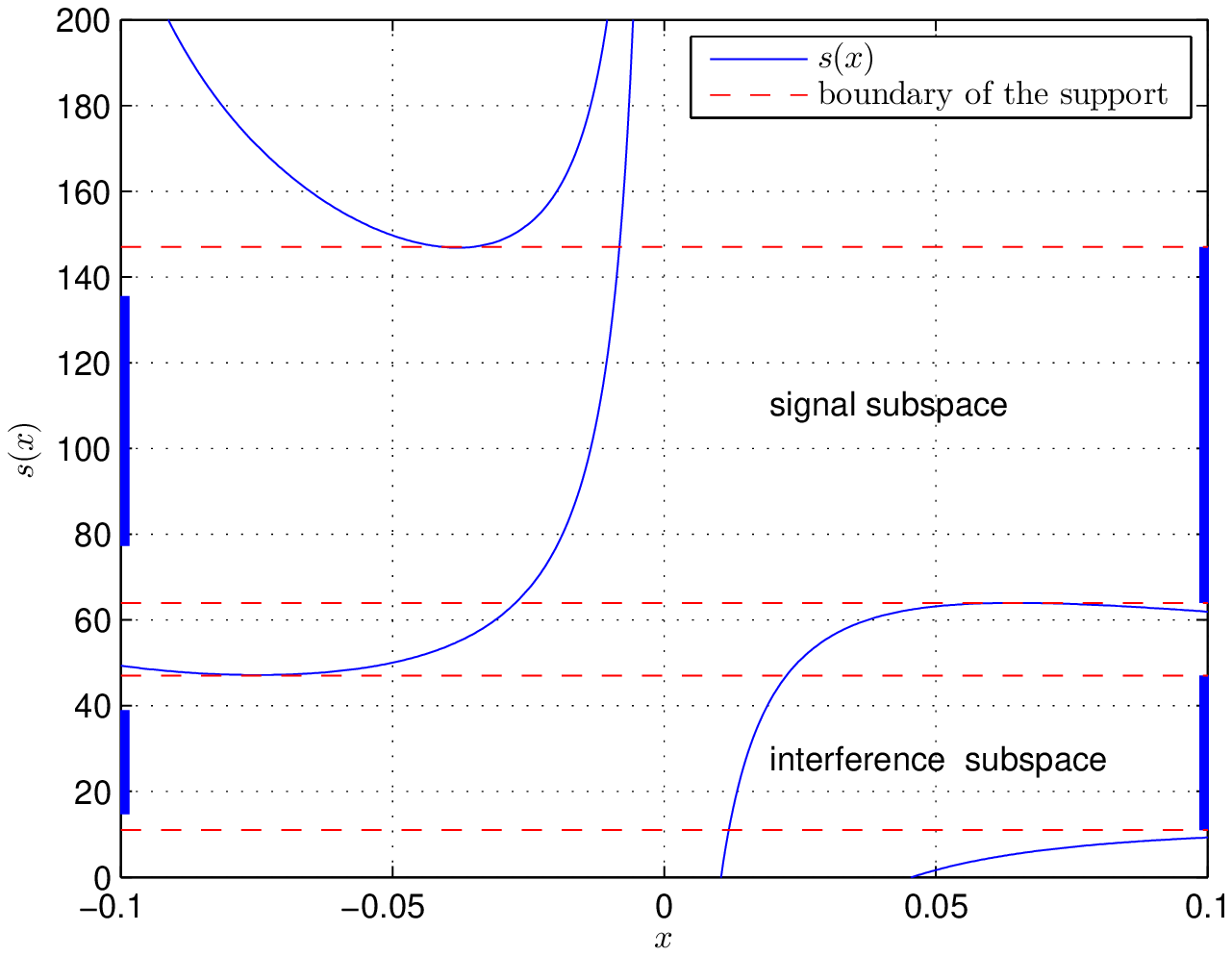}
\label{fig1}}
\hfil
\subfloat[approximate solution using~\eqref{eq:invfunction2}]{\includegraphics[width=2.7in]{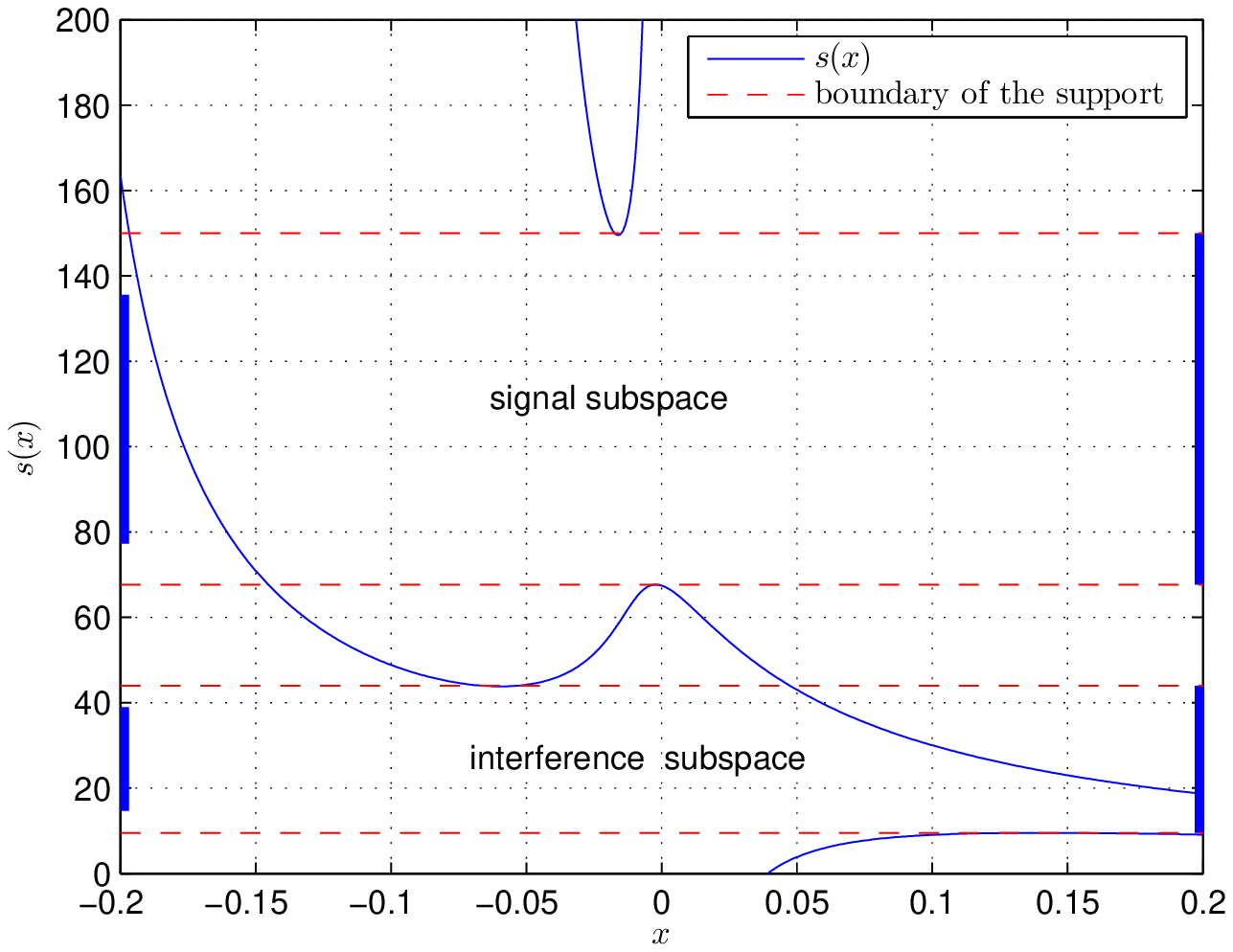}
\label{fig22}}
\caption{The nonzero support of the AED of ${Y_l Y_l^\dagger}/{M}$, $M=400$, $P=200$, $N=1000$, $K=5$, $L=4$, $P_s=-10 \mathrm{dB}$, $P_I=-16\mathrm{dB}$.}
\label{fig:supported boundary}
\end{figure*}

In Figs.\ref{fig1} and~\ref{fig22} we use~\eqref{eq:invfunction1} and~\eqref{eq:invfunction2}, respectively, to show the approximated boundaries of the eigenvalue distribution. The number of receive antennas $M=400$, $P=200$ AoAs, $L=4$ cells, $K=5$ users per cell, the coherence time $N=1000$ symbol periods\footnote{$N$ is taken to be a relatively large in order to highlight the effect of the number of AoAs and power ratio, rather than the effect of the number of measurements.}, the desired signal power $P_s=0.1 (-10 \mathrm{dB})$, whereas the interference power $P_I=0.025$ ($\approx -16\mathrm{dB}$), and high $\mathrm{SNR}$ is assumed (i.e., $\bm{W} = 0$). Further, the support of distribution of the desired signal and the interference can be read on the right vertical axes, while on the left vertical axes is the exact noise-free support obtained for an i.d. channel with $M=400$.

As expected, the eigenvalue clusters of the signal and the interference exhibit larger spread from both sides of the support, and hence the gap between the two clusters decreases significantly. In contrast, under the same setting, but i.d. channel, the eigenvalues show less spread and the gap between the two clusters is more pronounced. In practice, especially with noisy samples, it turns out that, realistic physical channel, relative to i.d. channel, incurs higher power ratio $P_s/P_I$ such that the spectrum splits into signal and interference eigenvalues in order to maintain a comparable BER. Otherwise, a high frequency reuse factor, for example, would be needed, which leads to poor use of spectrum resources

\begin{figure}[!ht]
\centering
\includegraphics[width=2.8in]{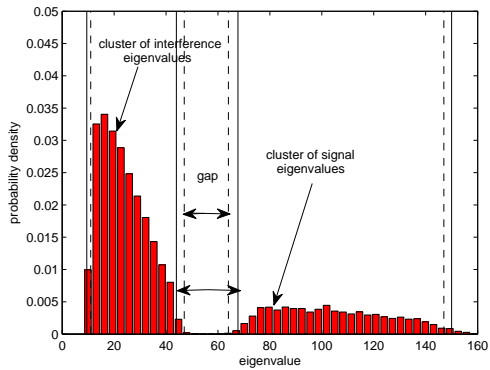}
\caption{Normalized histogram of nonzero eigenvalues of ${Y_l Y_l^\dagger}/{M}$, $M=400$, $P=200$, $N=1000$, $K=5$, $L=4$, $P_s=-10\mathrm{dB}$, $P_I=-16\mathrm{dB}$. The solid and dashed lines are the approximated boundaries obtained by~\eqref{eq:invfunction2} and~\eqref{eq:invfunction1}, respectively.}
\label{fig3}
\end{figure}

The supported boundaries computed by~\eqref{eq:invfunction1} and~\eqref{eq:invfunction2} are compared to the histogram of nonzero eigenvalues in Fig.\ref{fig3}. Obviously, the histogram is composed of two bulks of eigenvalues clustered around two eigenvalues, 25 and 100 (center eigenvalues). The solid and dashed vertical lines correspond to the endpoints approximated by~\eqref{eq:invfunction2} and~\eqref{eq:invfunction1}, respectively. Note that the approximated boundaries from~\eqref{eq:invfunction2} are almost in agreement with the boundaries of the two bulks obtained from the histogram of eigenvalues.

\begin{figure}[!ht]
\centering
\includegraphics[width=2.7in]{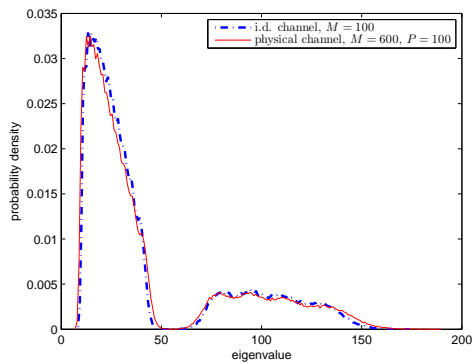}
\caption{Empirical pdfs of nonzero eigenvalues of ${Y_l Y_l^\dagger}/{M}$ for physical and i.d. channels, $N=1000$, $K=5$, $L=4$, $P_s=-10 \mathrm{dB}$, $P_I=-16\mathrm{dB}$.}
\label{fig4}
\end{figure}

Fig.\ref{fig4} shows the superimposed probability densities of eigenvalue for the physical channel (solid line) and the i.d. channel (dashed line). The parameters $N, K, L, P_s,$ and $P_I$ are the same as those in Fig.\ref{fig:supported boundary}. In the physical channel, $P=100$ identical AoAs for all users w.r.t. BS1, and the number of receive antennas $M$ is set to 600. On the other hand, the number of receive antennas in the case of i.d. channel is set to the number of AoAs, that is $M=100$. Note that the eigenvalue pdf of ${Y_l Y_l^\dagger}/{M}$ almost matches the eigenvalue pdf of the corresponding i.d. channel. Thus, this result implies that the gap between the two bulks of eigenvalues (corresponding to the signal subspace and the interference subspace) of the physical channel with $P$ AoAs cannot be further improved beyond what can be achieved when the channel is i.d. with $P$ receive antennas as $M$ increases unboundedly (see Cor.\ref{collary1111}). Nevertheless, array gain can be still retained as $M$ increases, but ultimately the performance will saturate as will be shown in the next subsection.

\begin{figure*}[!ht]
\centering
\subfloat[$P_1=P_2=P_3=P_4=200$]{\includegraphics[width=2.7in]{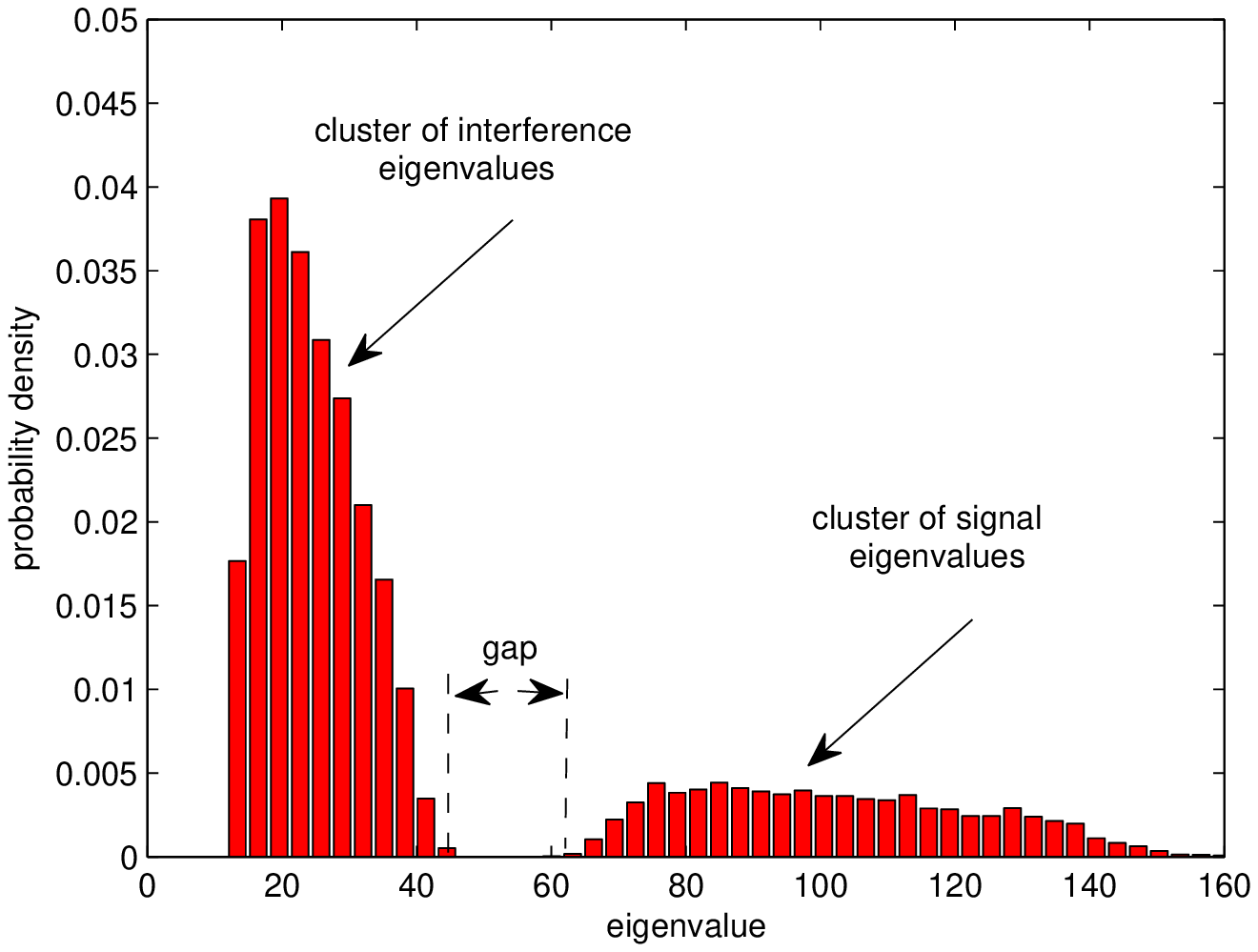}
\label{fig6}}
\hfil
\subfloat[$P_1=P_2=P_3=200$, $P_4=20$]{\includegraphics[width=2.7in]{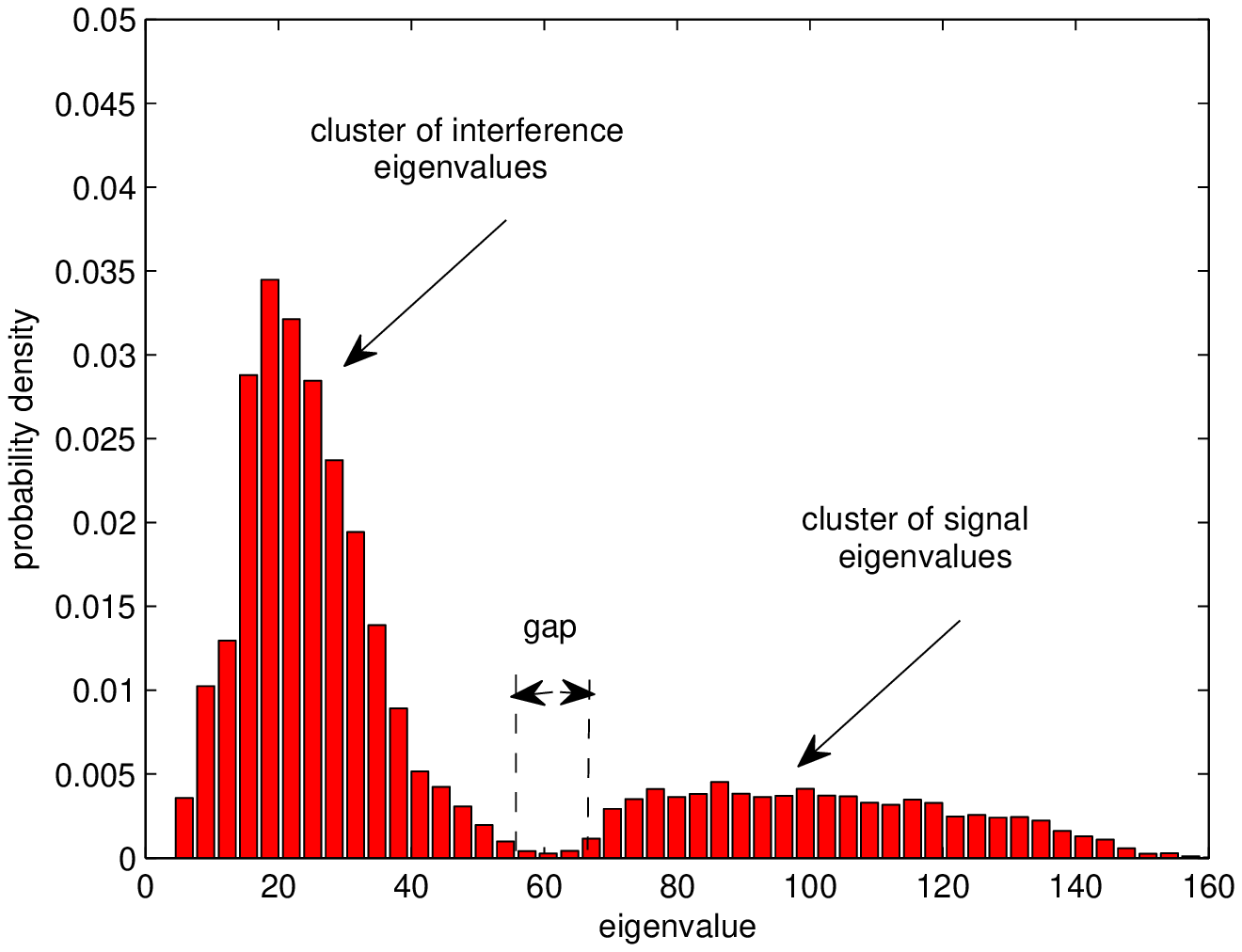}
\label{fig2}}
\caption{Normalized histogram of nonzero eigenvalues of ${Y_l Y_l^\dagger}/{M}$ with distinct AoAs, $M = 400$, $N=1000$, $K=5$, $L=4$, $P_s=-10 \mathrm{dB}$, $P_I=-16\mathrm{dB}$.}
\label{fig:Distinct AoA AED}
\end{figure*}

Finally, Figs.\ref{fig6} and~\ref{fig2} show the normalized histogram of eigenvalue distributions for physical channel model with distinct AoAs. The parameters $N, K, L, P_s,$ and $P_I$ are the same as those in Fig.\ref{fig:supported boundary}. In Fig.\ref{fig6} we independently generate four sets of equal number of AoAs ($P_1=P_2=P_3=P_4=200$) corresponding to cells 1, 2, 3, and 4 w.r.t. BS1. On the other hand, in Fig.\ref{fig2} we fix the number of AoAs of cells 1, 2, and 3 to 200 (i.e., a large number of AoAs), whereas the number of AoAs corresponding to cell 4 is set to 20 (i.e., very small number of AoAs). Compared with the histogram in Fig.\ref{fig6}, it is clear that the fourth cell renders the interference bulk of eigenvalues (left bulk) more wider and hence the gap is less pronounced.

\subsection{Bit error rate}

To give an intuitive feel of the effect of the physical channel on the performance of subspace method, in Fig.\ref{fig7} we show the uncoded BER versus the power ratio $P_I/P_S$. In our simulation we consider the uplink in a four-cell network with 5 users per cell and QPSK modulation scheme. The coherence time of the channel $N=400$ symbol periods, all signals are received from identical AoAs with cardinality $P = 200$ and per-user $\mathrm{SNR}=-5\mathrm{dB}$. To show the effect of increasing the number of receive antennas $M$, we use different values: $M=200, 400, 600$, where the antenna elements are critically-spaced. We consider full reuse of $K$ orthogonal pilot sequences across all cells. In addition, we use linear zero-forcing (ZF) for channel estimation (after applying SVD) and  matched filter (MF) for data detection. The performance comparison with the classical pilot-based channel estimator is also shown, in which ZF and MF receivers are used to estimate the channel and detect data, respectively.
\begin{figure}[!ht]
\centering
\includegraphics[width=2.7in]{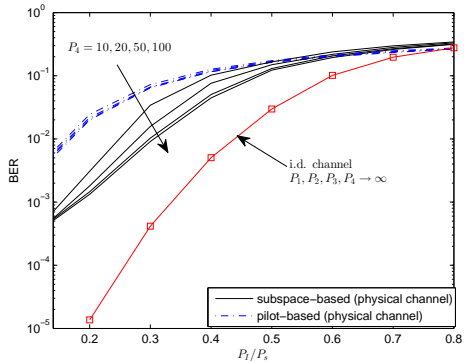}
\caption{ BER versus the power ratio for distinct AoAs, $P_1=P_2=P_3=100$, $M=400$, $K=5$, $L=4$, $N=400$, per-user $\mathrm{SNR}=-5\mathrm{dB}$, ZF is used for channel estimation and MF for data detection.}
\label{fig8}
\end{figure}

From Fig.\ref{fig7} we notice that the physical channel incurs loss of performance as opposed to the i.d. channel. For instance, when each BS is equipped with $400$ antennas and the number of AoAs is $200$, to achieve the same BER of $10^{-2}$, the power ratio should be roughly doubled, compared with the case of i.d. channel. Also, subspace-based scheme outperforms the pilot-based scheme in all cases except at the very high interference level, which is unlikely in practical scenarios. Further, the performance of the subspace-based scheme improves gradually with increasing $M$. As expected, its performance under this physical channel becomes closer to its performance under an i.d. channel with $M=200$ receive antennas. However, for very small interference levels, it further improves. Actually, this verifies the fact that when $M$ grows large, the eigenvalue distribution of this physical channel with $P$ AoAs becomes identical to that of an i.d. channel with $M=P$ receive antennas. It should be noted that it still benefits partly from array gain due to increasing $M$, and this array gain becomes more useful when the two subspaces start to detach (i.e., when two subspaces overlap, the estimation error of channel will get larger and thus have a dominant impact on the performance). After that point, the performance becomes dominated by the degrees of freedom of the channel only.

In Fig.\ref{fig8} we show the BER when distinct AoAs are used per each $K$ users in each cell w.r.t. BS1. To highlight the degradation of the performance due to the cell with the smallest number of AoAs, we fix the number of AoAs of cell 1, 2, and 3 to 100, while we vary the number of AoAs of the fourth cell, $P_4 = 10, 20, 50, 100$. It is clear that when the subspace-based scheme is used, the performance improves as $P_4$ increases, whereas it is almost the same for all values of $P_4$ when the pilot-based scheme is used. In subspace mehod, the gradual improvement of the performance can be interpreted as a consequence of gradual compression of the eigenvalues cluster of interference when $P_4$ increases (see Fig.\ref{fig6} and Fig.\ref{fig2}). However, when $P_4$ becomes sufficiently large, we observe a slight improvement of the performance (i.e., a saturation effect of the performance). This is because the performance becomes limited again by the other interfering cells associated with smaller number of AoAs. Further, it is expected that when $P_4$ decreases, the degradation in performance becomes even worse when the interfering power from the fourth cell is comparable to the power of the desired users.
\begin{figure}[!ht]
\centering
\includegraphics[width=2.7in]{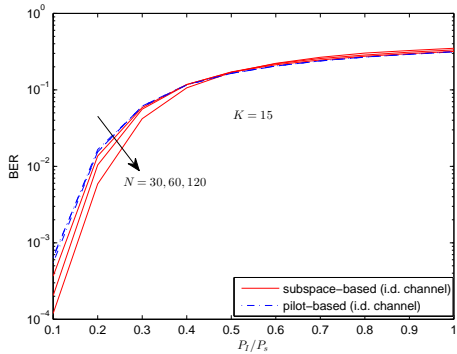}
\caption{BER when $N \sim KL$ under an i.d. channel, $M=400$, $K=15$, per-user $\mathrm{SNR}=0 \mathrm{dB}$, $L=4$, ZF is used for channel estimation and MF for data detection. }
\label{fig9}
\end{figure}

Finally, in Fig.\ref{fig9} we simulate the BER when the number of users of the network is comparable to the coherence time of the channel. In these scenarios the all-orthogonal pilot-based scheme doesn't work. Instead, we compare its performance with the classical pilot-based scheme. We assume an i.d. channel, $K=15$, $L=4$ and the coherence time $N$ varies around $KL=60$, i.e., $N=30, 60, 120$. It can be observed from Fig.\ref{fig9} that in all scenarios, especially the critical cases when $N=30$ and $N=60$, the subspace-based scheme outperforms the pilot-based scheme. Its performance also improves with increasing $N$ while the performance of pilot-based scheme is almost the same for all values of $N$. This means that subspace-based scheme exhibits better flexibility than the classical pilot-based scheme and all-orthogonal pilot-based scheme.

\section{Conclusion}\label{sec6}
Pilot contamination is usually considered as a performance bottleneck in massive MIMO systems, if the conventional linear channel estimation method is used. As a departure from the linear estimation framework, the subspace-based scheme proposed in~\cite{blind_pilot_cont} exhibits better flexibility under various scenarios. In addition, it offers the potential of completely eliminating the pilot contamination effect. However, previous works are based on the common assumption of independent channels, which would be violated by real-world channels where the number of scatterers and AoAs might be both limited.

For the more sensible physical channel model, in this paper, we derived approximate analytic expressions in Stieltjes domain for the corresponding eigenvalue distributions in both scenarios of \emph{identical} and \emph{distinct} AoAs among the cells. The results demonstrate that the physical channel will exhibit a larger spectral spread, thus resulting in a smaller gap between the two eigenvalue clusters, which correspond to signal and interference subspaces. Therefore, to obtain the same performance as of the i.d. channels, we must guarantee a larger power ratio between the intended users and the interfering users. In particular, for the scenarios of distinct number of AoAs, the required power difference is determined by the cell with the smallest number of AoAs. Moreover, it is shown that adding more antennas does not significantly affect the above spectral gap, even though, the array gain could be reaped until saturation.

For future research, since the blind subspace-based channel
estimation scheme relies primarily on the eigenvalues spread
of the channel (non-Bayesian or data-driven method), rather
than the statistical properties of the channel (Bayesian method), the
exploration of the possibility of combining the two methods
is of importance from the perspective of system performance,
especially in the scenario of bounded angular spreads.

Finally, the results in this paper demonstrate the significant impact of the physical channel with finite AoAs on the performance of massive MIMO system, especially for the blind subspace-based channel estimation methods.

\appendix[discussion on the freeness condition]
In this appendix, we prove the asymptotic freeness of the products of matrices in~\eqref{final marix product} and~\eqref{eq:SJG2prime}. Let $\{\bm{E}_i\}\in \mathcal{C}^{v\times v}$ and $\{\tilde{\bm{E}}_{\acute{i} } \} \in \mathcal{C}^{v\times v}$ be two families of bi-unitarily invariant\footnote{A bi-unitarily invariant matrix is a matrix whose the joint distribution of its entries is invariant when both left- and right- multiplied by unitary matrices.} random matrices, whose AED's, as $v\to \infty$, converge almost surely to non-random distributions with compacted supports. Further, let $\{\bm{Z}_j\} \in \mathcal{C}^{v\times v}$ and $\{\tilde{\bm{Z}}_{\acute{j} }\} \in \mathcal{C}^{v\times v}$ be two families of non-random diagonal matrices with almost sure convergence of their AED's to non-random distributions with bounded eigenvalues as $v\to \infty$. Then from \cite[Thm. 4.3.11]{Hiai:2006:SLF:1204180}, the family
$\{ \{ \bm{E}_i \}, \{ \tilde{\bm{E}}_{\acute{i} }^\dagger \}, \{ \bm{Z}_j \}, \{ \tilde{\bm{Z}}_{\acute{j} }^\dagger \} \}, \forall i, \acute{i}  \in I, j, \acute{j} \in J $
is asymptotically free. It is important to note that it is not necessary for $\bm{E}_{i}$ and $\tilde{\bm{E}}_{\acute{i}}$ to be square matrices. This can be seen from the fact that $\bm{Z}_{j}$ and $\tilde{\bm{Z}}_{\acute{j}}$ can have different dimensions, e.g., an arbitrary number of the last entries can be zeros. Equivalently, $\bm{E}_{i}$ and $\tilde{\bm{E}}_{\acute{i}}$ can be considered as non-square matrices whereas $\bm{Z}_{j}$  and $\tilde{\bm{Z}}_{\acute{j}}$ are two square matrices, such that the combination of all matrices is defined.

Now, let us take the three cases: $\bm{E}_{i}=\tilde{\bm{E}}_{\acute{i}}$ and $\bm{Z}_{j}={\tilde{\bm{Z}}}_{{\acute{j}}}=\bm{I}$; $\bm{E}_{i}=\tilde{\bm{E}}_{{\acute{i}}}$ and ${\tilde{\bm{Z}}}_{{\acute{j}}}=\bm{I}$; $\bm{E}_{i}=\tilde{\bm{E}}_{{\acute{i}}}$. Then for each case we define the respective products:
\begin{eqnarray}
\bm{E}_{i} \bm{E}_{i}^\dagger, &i\in I\label{eq:app1} \\
\bm{E}_{i}^\dagger (\bm{U}^\dagger \bm{Z}_{j}\bm{U} )\bm{E}_{i}, &i\in I,j\in J\label{eq:app2}\\
\bm{Z}_{j} \bm{E}_{i} \bm{E}_{i}^\dagger \tilde{\bm{Z}}_{\acute{j}}^\dagger, &i\in I,j,\acute{j}\in J\label{eq:app3}
\end{eqnarray}
where the factor $(\bm{U} \bm{Z}_i \bm{U}^\dagger )$ in~\eqref{eq:app2} follows from the fact that $\bm{E}_{i}$ is bi-unitarily invariant and hence can be replaced by $\bm{U}\bm{E}_{i}\bm{V}^\dagger$, where $\bm{U}$ and $\bm{V}$ are unitary matrices. Note that $\bm{U}^\dagger \bm{Z}_{j} \bm{U}$ is a unitarily invariant Hermitian matrix with AED converges to a non-random distribution (i.e., the eigenvectors are distributed in a maximally random way implying that $\bm{V}^\dagger \bm{Z}_j \bm{V}$ fulfills the condition of~\cite[Thm. 4.3.11]{Hiai:2006:SLF:1204180}), and by~\cite[Thm. 4.3.11]{Hiai:2006:SLF:1204180} any subset of~\eqref{eq:app1},~\eqref{eq:app2} and~\eqref{eq:app3} forms a family that is asymptotically free.

Now, from~\eqref{final marix product}, since the entries of $\bm{B}_k$ and $\bm{C}_k$ are assumed Gaussian with zero mean and unit variance, then they form  bi-unitarily invariant random matrices each with the AED  converges to the well-known Mar\v{c}enko-Pasture law~\cite{Marcenko:1967}. Then the Hermitian matrices $(\bm{A}_k\bm{B}_k)^\dagger (\bm{A}_k\bm{B}_k)=\bm{B}_k^\dagger (\bm{A}_k^\dagger \bm{A}_k)\bm{B}_k$ and $\bm{C}_k \bm{C}_k^\dagger$ correspond, respectively, to~\eqref{eq:app2} and~\eqref{eq:app1}. Further, because the assumption of i.i.d. of $\bm{A}_k^\dagger \bm{A}_k$ incurs insignificant penalty compared with the Vandermonde matrix, it follows from~\eqref{eq:app1} and~\eqref{eq:app2} that $\bm{B}_k^\dagger$, $\bm{A}_k^\dagger \bm{A}_k$, $\bm{B}_k$, and $\bm{C}_k \bm{C}_k^\dagger$ form a family that is free asymptotically.

Finally, the product $\tilde{\bm{D}}^{1/2}{\bm{X}}{\bm{X}}^\dagger \tilde{\bm{D}}^{1/2} (\bm {S} \tilde{\bm{H}} )^\dagger  (\bm {S} \tilde{\bm{H}}) $ in~\eqref{eq:SJG2prime} can be treated in the same way. Note that $\tilde{\bm{D}}^{1/2}{\bm{X}}{\bm{X}}^\dagger \tilde{\bm{D}}^{1/2}$ and $(\bm {S} \tilde{\bm{H}} )^\dagger  (\bm {S} \tilde{\bm{H}})$ are special instances of~\eqref{eq:app3} and~\eqref{eq:app2}, respectively. Therefore, the asymptotic freeness property holds true for the products in~\eqref{eq:SJG2prime}.

\section*{Acknowledgment}
The authors are indebted to the anonymous reviewers for several helpful comments and suggestions. This work is supported by Natural Science Foundation of China under grant 61231007 as well as the National Science and Technology Major Project under
grant 2013ZX03001015-002, and Dr. Gesbert acknowledges the partial support of European project HARP under the FP7 ICYT Objective 1.1. 

\bibliographystyle{IEEEtran}
\bibliography{IEEEabrv,references}

\end{document}